\documentclass[3p]{elsarticle}
\biboptions{sort&compress}

\usepackage[utf8x]{inputenc}
\usepackage[cmex10]{amsmath}
\usepackage{amssymb}
\usepackage{color}
\usepackage{setspace}
\usepackage{graphicx}
\usepackage{geometry}
\usepackage{multirow}
\usepackage{multicol}
\usepackage[ruled]{algorithm}
\usepackage{algpseudocode}
\usepackage{fixltx2e}
\usepackage{placeins}
\usepackage{xspace}
\usepackage{lineno}
\usepackage{pxfonts}
\usepackage{float}
\usepackage{subcaption}
\usepackage{booktabs}
\usepackage{paralist}
\usepackage{xcolor}
\usepackage[normalem]{ulem}

\usepackage{enumitem}
\setlist[enumerate]{itemsep=-0.25\baselineskip}
\setlist[itemize]{itemsep=-0.25\baselineskip}

\usepackage{xr}
\definecolor{darkgreen}{rgb}{0,.55,.4}
\definecolor{dgreen}{rgb}{0,0.6,0.2}
\definecolor{gray}{rgb}{0.5,0.5,0.5}
\definecolor{dark}{rgb}{0.4,0.4,0.4}
\definecolor{silver}{rgb}{0.85,0.85,0.85}
\definecolor{mauve}{rgb}{0.58,0,0.82}
\definecolor{hlite}{rgb}{1,1,.75}
\definecolor{cyan}{rgb}{0,.3,0.8}
\definecolor{purple}{rgb}{.4,0,0.9}

\usepackage{listings}
\MakeRobust{\Call}
\algnewcommand{\LComment}[1]{\State \(\triangleright\) {\color{black} \textbf{#1}}}
\newcommand{\mpicommworld}{\texttt{MPI\_COMM\_WORLD}\xspace}
\newcommand{\USERMESO}{\texttt{\textsubscript{\textit{USER}}MESO}\xspace}
\newcommand{\code}[1]{\textbf{{\texttt{#1}}}}
\newcommand{\hlcode}[1]{\colorbox{hlite}{#1}}
\newcommand{\revised}[1]{#1} 
\newcommand{\replace}[2]{#2} 

\journal{Journal of Computational Physics}

\begin{document}

\begin{frontmatter}

\title{Multiscale Universal Interface: A Concurrent Framework for Coupling Heterogeneous Solvers}

\author[brown]{Yu-Hang Tang}
\ead{yuhang\_tang@brown.edu}

\author[kobe]{Shuhei Kudo}
\ead{138x211x@stu.kobe-u.ac.jp}

\author[brown]{Xin Bian}
\ead{xin\_bian@brown.edu}

\author[brown]{Zhen Li}
\ead{zhen\_li@brown.edu}

\author[brown,pnnl]{George Em Karniadakis\corref{cor1}}
\ead{george\_karniadakis@brown.edu}

\cortext[cor1]{Corresponding author}
\address[brown]{Division of Applied Mathematics, Brown University, Providence, Rhode Island, USA}
\address[kobe]{Graduate School of System Informatics, Kobe University, 1-1 Rokkodai-cho, Nada-ku, Kobe, 657-8501, Japan}
\address[pnnl]{Collaboratory on Mathematics for Mesoscopic Modeling of Materials, Pacific Northwest National Laboratory, Richland, WA 99354, USA}

\begin{abstract}
Concurrently coupled numerical simulations using heterogeneous solvers are powerful tools for modeling multiscale phenomena. However, major modifications to existing codes are often required to enable such simulations, posing significant difficulties in practice. In this paper we present a C++ library, \textit{i.e.} the Multiscale Universal Interface (MUI), which is capable of facilitating the coupling effort for a wide range of multiscale simulations. The library adopts a header-only form with minimal external dependency and hence can be easily dropped into existing codes. A \textit{data sampler} concept is introduced, combined with a hybrid dynamic/static typing mechanism, to create an easily customizable framework for solver-independent data interpretation. The library integrates MPI MPMD support and an asynchronous communication protocol to handle inter-solver information exchange irrespective of the solvers' own MPI awareness. Template metaprogramming is heavily employed to simultaneously improve runtime performance and code flexibility. We validated the library by solving three different multiscale problems, which also serve to demonstrate the flexibility of the framework in handling heterogeneous models and solvers. In the first example, a Couette flow was simulated using two concurrently coupled Smoothed Particle Hydrodynamics (SPH) simulations of different spatial resolutions. In the second example, we coupled the deterministic SPH method with the stochastic Dissipative Particle Dynamics (DPD) method to study the effect of surface grafting on the hydrodynamics properties on the surface. In the third example, we consider conjugate heat transfer between a solid domain and a fluid domain by coupling the particle-based energy-conserving DPD (eDPD) method with the Finite Element Method (FEM).
\end{abstract}


\end{frontmatter}

\let\thefootnote\relax\footnote{The authors wish it to be known that the first two authors each contributed indispensably in this interdisciplinary work and thus, in their opinion, should be regarded as co-first authors.}
\let\thefootnote\relax\footnote{The Multiscale Universal Interface library is freely available under the GPLv3 license at \texttt{http://www.cfm.brown.edu/repo/release/MUI/}.}



\section{Glossary}

A \textbf{solver} is a computer program that can carry out a given type of numerical simulation. It may execute in the single-program-multiple-data (SPMD) mode for parallelization. The same solver can be invoked multiple times separately during a certain simulation.

A \textbf{simulation} is the act of using one or multiple solvers to perform a numerical modeling task. A \textbf{system} is the entire set of physical time-space involved in a simulation. A \textbf{subdomain}, or simply \textbf{domain}, is a subset of a system that is handled by a single solver. The numerical modeling result on a domain may depend on information from other subdomains of the system and vice versa.

An \textbf{interface} is a communication layer for exchanging information such as the boundary conditions between two or more solvers. A \textbf{sender} is a solver process which is pushing information into an interface, while a \textbf{receiver} is pulling information from the interface. A solver can be both sender and receiver at the same time. \textbf{Peers} are the collection of receivers with regard to a given sender. A \textbf{MUI interface} is an instance of our software interface enabling the physical interfaces.


\section{Introduction}
\label{sec:intro}

The potential of multiscale modeling lies in its ability of probing properties of hierarchical systems by capturing events that occur across a wide range of time and length scales that exceed the capability of any single solver and method~\revised{\cite{praprotnik2008multiscale,mohamed2010review,weinan2011principles}}. More recently, one specific branch, \textit{i.e.} domain decomposition-based concurrent coupling, has seen rapid development since it allows on-the-fly information exchange and interaction between multiple simulation subdomains handled by different solvers.

Multiscale concurrent coupling using domain decomposition dates back to the classical Schwartz alternating method~\cite{lions1988schwarz,xu1992iterative}, where solutions of a partial differential equation (PDE) on two subdomains can be pursued iteratively. In this method, the calculation for subdomain A is first performed with an enforced pseudo-boundary which extends into the other subdomain B, while the solutions on the pseudo-boundary are dictated to be the known values of B at corresponding locations. A similar calculation is then performed for subdomain B. This procedure is repeated iteratively until convergence of solution in the hybrid region or global domain is achieved. In general, there are two practical strategies to enforce the pseudo-boundary between two subdomains. The first is state-variable, \textit{e.g.} density, velocity, etc., based coupling, where the constraints are placed on the state variables on the two pseudo-boundaries \replace{alternatively}{alternately} \revised{\cite{walther2012multiscale,chen2014particle}}. The other is flux based coupling, where the flux, \textit{e.g.} mass flux, momentum flux, etc., flowing into/out of one subdomain is compensated by the other subdomain so that the laws of conservation are respected \revised{\cite{delgado2008concurrent,patronis2014multiscale}}.

\replace{However}{Despite existing theoretical developments}, implementing concurrently coupled simulations remains difficult. On one hand, hard-coding remains commonplace in projects that employ \textit{ad hoc} \revised{coupling} approaches. Such practice can quickly become an obstacle when further development is needed. On the other hand, despite the richness of available coupling schemes and tools \cite{Allan2002CCA,Lefantzi2003CCA,mcinnes2006PDE, uintah2000,Parker2006Uintah,Berzins2010uintah}, adapting existing code to meet the programming interface specification of a coupling framework frequently leads to code refactoring that consumes a substantial amount of man-hours~\cite{allan2004odepack}. Such frameworks could also be cumbersome, especially for theorists without an expertise in software engineering, when prototyping new coupling schemes.

As far as we know, a general and non-expert-friendly software library that assists the  concurrent coupling of independently developed solvers remains unavailable. The Multiscale Universal Interface (MUI) project aims to fill in this gap by creating a light weight plugin library that can glue together essentially all numerical methods including, but not limited to, Finite Difference, Finite Volume, Finite Element, Spectral Method, Spectral Element Method, Lattice Boltzmann Method, Molecular Dynamics, Dissipative Particle Dynamics and Smoothed Particle Hydrodynamics. Hence, it can deal with Lagrangian or Eulerian descriptions or a mixture of both. It is expected to be able to accommodate a wide range of coupling schemes regardless of the quantities being exchanged, the equations being solved, the time stepping pattern and/or the degree of spatial and temporal separation. 

In order to achieve such a high level of universality, MUI is designed to avoid defining the math behind the coupling procedure, \textit{i.e.} it does not specify \textit{which} and \textit{how} quantities are coupled. Instead, it provides services to facilitate the effort of constructing arbitrary coupling schemes by enabling the communication and interpretation of arbitrary physical quantities using arbitrary data types as demanded by each participating solver.

MUI is simple to use, in the sense that existing solvers do not have to be refactored before using it. MUI provides a very small set of programming interfaces instead of dictating any from the solver. The entire library is coded in a header-only fashion with the Message Passing Interface (MPI) being the only external dependency. Hence, it can be used in exactly the same way as the C++ standard library without pre-compilation. It does not interfere with existing intra-solver communications for solvers using MPI.

MUI is also fast, in the sense that using it only consumes a small amount of CPU time as compared to that used by the solver itself. To achieve this goal we heavily employ the C++ generic programming/template metaprogramming feature to eliminate the abstraction overhead that may otherwise arise when maintaining the high-level flexibility of the framework.


\section{Data Interpretation Framework}

\subsection{Usage Overview}
\label{sec:uover}

\begin{figure}
\centering
\includegraphics[width=3.5in]{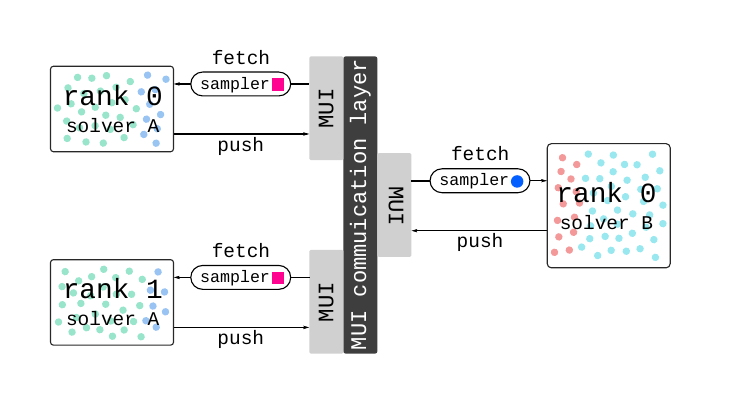}
\caption{MUI facilitates the exchange of information across solvers by letting solvers push and fetch data points. Flexibility of interpolation is achieved by allowing the expression of interpolation algorithms as samplers as well as by accepting data points of arbitrary types.}
\label{fig:usagepattern}
\end{figure}

\revised{
As visualized in Figure~\ref{fig:usagepattern}, MUI assumes a \code{push}-\code{fetch} workflow and serves as the data exchange and interpretation layer between solvers. While more concrete examples on the usage of MUI are given in Section~\ref{sec:benchmark}, the list below outlines the typical steps for incorporating MUI into an existing solver:
\begin{enumerate}
  \item Substituting the MPI global communicator (Section~\ref{sec:MPIMPMD});
  \item Allocating MUI objects;
  \item Identifying code regions that supply information to peer solvers and pushing the data as points using MUI (Section~\ref{sec:datapoints});
  \item Identifying code regions that require information from peer solvers and fetching through MUI's sampling interface (Section~\ref{sec:datasampler});
  \item Configuring inter-solver synchronization (Section~\ref{sec:timeframes});
  \item Optimizing performance by managing memory allocation, simplifying communication topology and tuning low-level traits \textit{etc.} (Section~\ref{sec:timeframes}, Section~\ref{sec:smartsend}, and Section~\ref{sec:custom}). This step is not mandatory for obtaining correct results but may have an impact on simulation efficiency.
\end{enumerate}

}

\subsection{Data Points}
\label{sec:datapoints}

\begin{figure}
\centering
\includegraphics[width=3.5in]{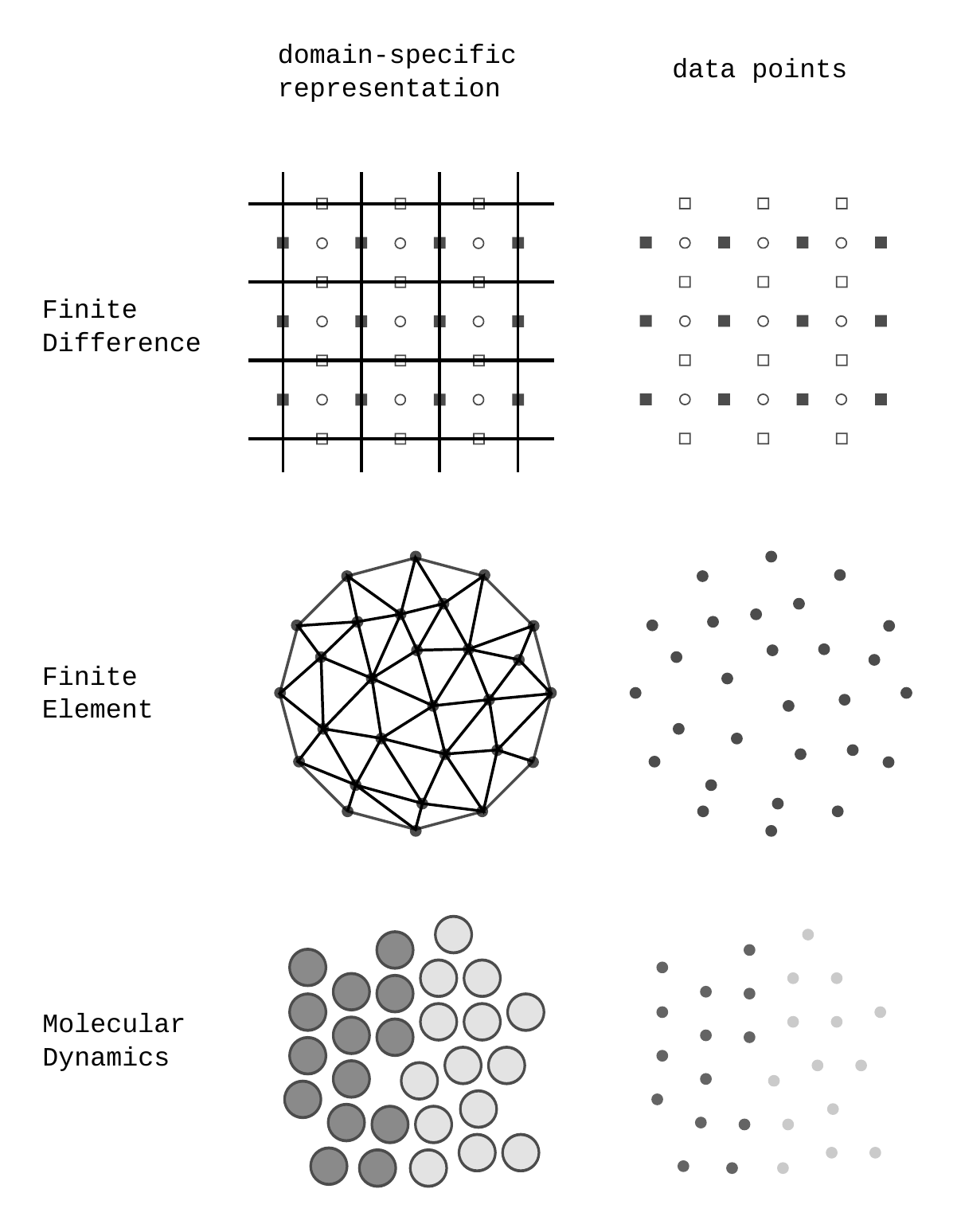}
\caption{Any discrete systems can be generalized as a cloud of data points by ignoring the domain-specific knowledge such as meshes.}
\label{fig:data_point}
\end{figure}

A universal coupling framework entails a generalized data representation framework. By observing the fact that discretization is the first step toward any numerical approximation, we realize that essentially every simulation system can be treated as a cloud of \textit{data points} each carrying three attributes, \textit{i.e.} position, type, and value, as shown in Fig. \ref{fig:data_point}. The points might be arranged on a regular grid or connected by a certain topology in some of the methods, but for the sake of generality it is useful to ignore this information temporarily.

MUI defines a generic \code{push} method for solvers to exchange points carrying different types of data in a unified fashion. The method assumes the signature:
\begin{lstlisting}
template<typename TYPE> inline
bool push( std::string name, mui::point location, TYPE value );
\end{lstlisting}

The \code{push} method can accept data points of arbitrary type because it takes the type of the value as a template argument. Points belonging to the same physical variable, \textit{i.e.} points pushed under the same name, are accumulated in a continuous container and sent out to receivers collectively to avoid fragmented communication. Note that the solver takes the responsibility of determining which points get pushed in, because the determination of the interface region uses mostly \textit{prior} knowledge and hence it does not necessarily require direct aid from the coupling library.

\subsection{Data Sampler}
\label{sec:datasampler}

\begin{figure}
  \centering
  \begin{subfigure}[t]{0.48\columnwidth}
    \centering
    \includegraphics[width=1.75in]{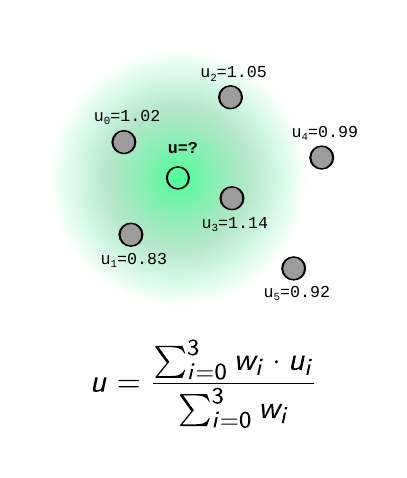}
    \caption{Weighted average}
    \label{fig:data_sampler_b}
  \end{subfigure}
  \begin{subfigure}[t]{0.48\columnwidth}
    \centering
    \includegraphics[width=1.75in]{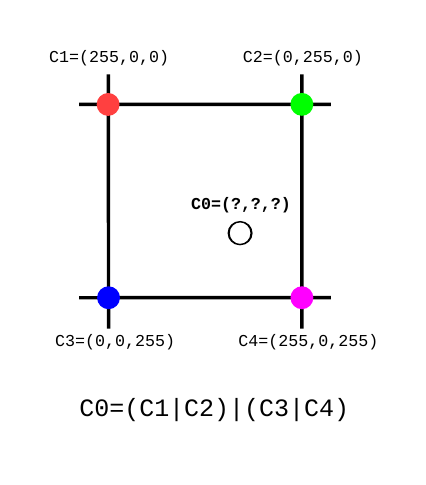}
    \caption{Bilinear color interpolation}
    \label{fig:data_sampler_a}
  \end{subfigure}
  \caption{Weighted kernel sampling and Texture sampling.}
  \label{fig:data_sampler}
\end{figure}

A generic \code{fetch} method for universal data interpretation on top of the data point representation is less trivial, however. The challenge of achieving universality here lies in the fact that solvers may be agnostic of the math and method used by their peers. Thus, a finite difference code might find itself in need of the value of pressure at grid point $(x_0,y_0)$, yet none of the vertices supplied by its peer finite element solver lies exactly on that point. In this case, the MUI interface is not supposed to simply throw out an exception. Possible solutions could be to use the pressure value defined at the nearest point, or to perform some sort of weighted interpolation using nearby points as shown in \revised{Figure~\ref{fig:data_sampler_b}}. The decision for the best algorithm requires knowledge beyond the reach of MUI, but we implemented a flexible data interpretation engine so that users can choose to plug in an appropriate one with minimal effort.

Such engine is enabled by the \textit{data sampler} construct, which is derived from the concept of texture sampling in computer graphics~\cite{hughes2013computer}. A texture is essentially a rasterized image of discrete pixels being mapped onto some 3D surface. As a result of 3D projection and transformation, there is no one-to-one correspondence between the pixels of the surface as shown on the screen and the pixels on the texture, and the color of the texture at a fractional coordinate could be displayed. In this case, as shown in Fig. \ref{fig:data_sampler_a}, the graphics hardware performs an interpolation (usually bilinear) of the pixel values adjacent to the requested fractional coordinate, and return the interpolated value as the color at the requested point.

The MUI data sampler works in a similar, but enhanced, way: each physical quantity is treated as a \textit{samplable object}, while data samplers are used to interpolate values from the cloud of discrete data points contained in it. A MUI sampler is a class implementing the interfaces \code{filter} and \code{support} as shown by the example of the MUI built-in Gaussian kernel sampler in Listing~\ref{list:samplercode}. A line-by-line explanation of the C++ code is given below:
\begin{itemize}
  \item line 1: class template argument declaration. By making the input and output type of samplers template arguments, it is possible to reuse the same interpolating algorithm without duplicating the code merely for the different data types used in different solvers.
  \item line 4-6: The internal basic data types of MUI are globally parameterized in a configuration class as detailed in Section~\ref{sec:custom}.
  \item line 8-11: Constructor that sets up the shape parameters of the Gaussian kernel.
  \item line 13-27: The \code{filter} method performs data interpolation/interpretation using data points fed by MUI. The MUI virtual container object maps to the subset of the data points that falls within the sampler's support while its usage pattern resembles that of \code{std::vector}.
  \item line 29-31: MUI uses geometry information provided by the \code{support} method as the extent of the sampler's support to efficiently screened off outlying particles with an automatically tuned spatial searching algorithm.
  \item line 33-36: storage for sampler parameters, etc.
\end{itemize}

\begin{lstlisting}[caption=The implementation of the MUI built-in Gaussian kernel sampler, label=list:samplercode]
template<typename OTYPE, typename ITYPE=OTYPE, typename CONFIG=default_config>
class sampler_gauss {
public:
	using REAL       = typename CONFIG::REAL;
	using INT        = typename CONFIG::INT;
	using point_type = typename CONFIG::point_type;

	sampler_gauss( REAL r_, REAL h_ ) :
	  r(r_),
	  h(h_),
	  nh(std::pow(2*PI*h,-0.5*CONFIG::D)) {}

	template<template<typename,typename> class CONTAINER>
	OTYPE filter( point_type focus, const CONTAINER<ITYPE,CONFIG> &data_points ) const {
		REAL  wsum = 0;
		OTYPE vsum = 0;
		for(INT i = 0 ; i < data_points.size() ; i++) {
			auto d = (focus-data_points[i].first).normsq();
			if ( d < r*r ) {
				REAL w = nh * std::exp( (-0.5/h) * d );
				vsum += data_points[i].second * w;
				wsum += w;
			}
		}
		if ( wsum ) return vsum / wsum;
		else return OTYPE(0);
	}

	geometry::any_shape<CONFIG> support( point_type focus ) const {
		return geometry::sphere<CONFIG>( focus, r );
	}

protected:
	REAL r;
	REAL h;
	REAL nh;
};
\end{lstlisting}

The sampling procedure works as:
\begin{enumerate}
  \item Solver invokes the \code{fetch} method of MUI with a \code{point of interest} and a sampler;
  \item MUI collects all points that lies within the sampler's support around the point of interest into a virtual container;
  \item MUI feeds the sampler with the collected points and lets the sampler perform its own interpolation;
  \item The sampler returns the interpolation result back to the user/solver through MUI.
\end{enumerate}

MUI achieves generality in interpolation by allowing users to easily create new samplers to express custom approximation algorithm that can leverage domain-specific knowledge of the system. The value at an arbitrary desired location can be obtained by using samplers that interpolate values from nearby points. In addition, a single piece of sampler code can be used for different data types, \textit{e.g.} \code{float}, \code{double} or \code{int}, because the \code{filter} and \code{fetch} method take the type of the data points as a template argument.

The sampling framework makes it possible for users to fully focus on the design of algorithms while delegating the data management job to MUI. To further simplify the usage, MUI includes several predefined samplers such as a Gaussian kernel sampler, a nearest neighbor sampler, a moving average sampler, an exact point sampler, \textit{etc}.

\subsection{Typing system}
\label{sec:typing}

MUI implements a hybrid dynamic/static typing system to combine the performance of static typing with the flexibility of dynamic typing. The dynamic typing behavior of MUI is performed at the level of physical quantities. The value of the first data point received by the \code{push} method for each physical quantity determines the type of the quantity, while the type of subsequently pushed data points will be examined against the type of the existing storage object. The entire storage object is type-dispatched only once on the receiver's side for each sampling request. Hence, MUI does not have to perform the expensive type-dispatching for each data points. This coarse-grained dynamic-typing technique is especially important for sampling where data points are being frequently accessed.

The system uses a type list to enumerate all possible types that may be handled by MUI and to automatically generate type-dispatching code. A type list is essentially an instantiation of \replace{the}{a} variadic class template \replace{\code{mui::tuple\_list\_t}}{} with the template arguments being the list members. Using recursive templates we can either query the type of a list member using an index (a compile-time constant) or check the index of a type in a given list. A default type list containing frequently used C++ built-in data types is predefined in MUI's default configuration.

Support for new types can be trivially added into MUI by \textbf{1)} adding the type into the predefined type list; and \textbf{2)} defining the insertion and extraction operator of the type with regard to \replace{\code{mui::istream }and \code{mui::ostream}. \code{mui::istream} and \code{mui::ostream} are}{} the data serialization classes in MUI \replace{sharing}{which share} the same usage pattern with \replace{\code{std::iostream}}{the C++ standard input and output streams}. Since MUI predefines the insertion and extraction operators for all C++ primitive types, an overloaded operator for any composite type can be implemented easily in terms of the primitive ones as illustrated in Listing~\ref{list:overload}.

\begin{lstlisting}[caption=The insertion and extraction operator for type \code{bond} can be overloaded in a straightforward way as shown below. \revised{\code{mui::istream} and \code{mui::ostream} are the built-in data serialization classes in MUI.}, label=list:overload]
// bond is a composite data type
struct bond {
  // double is a primitive data type
  double k, r0;
};

// overload serialization operator
mui::ostream& operator <<( mui::ostream& ost, const bond &v ) {
  // enumerate over primitive members
  return ost << v.k << v.r0;
}

// overload deserialization operator
mui::istream& operator >>( mui::istream& ist, bond &v ) {
  return ist >> v.k >> v.r0;
}
\end{lstlisting}

\subsection{Storage and Time coherence}
\label{sec:timeframes}

Regardless of the actual simulation algorithm, the main body of a solver is essentially a time marching loop in which the quantity of interest is being iteratively solved. Hence, points of the same quantity may be sent to a MUI interface repeatedly during a simulation. However, it is inevitable that one solver may run faster than its peer due to factors such as intrinsic performance disparity, load imbalance and transient interruption. In such situation\revised{s}, data points from a later time step may override previous ones belonging to the same quantity before the receiver could ever get a change to sample them. 

\begin{figure}
\centering
\includegraphics[width=3in]{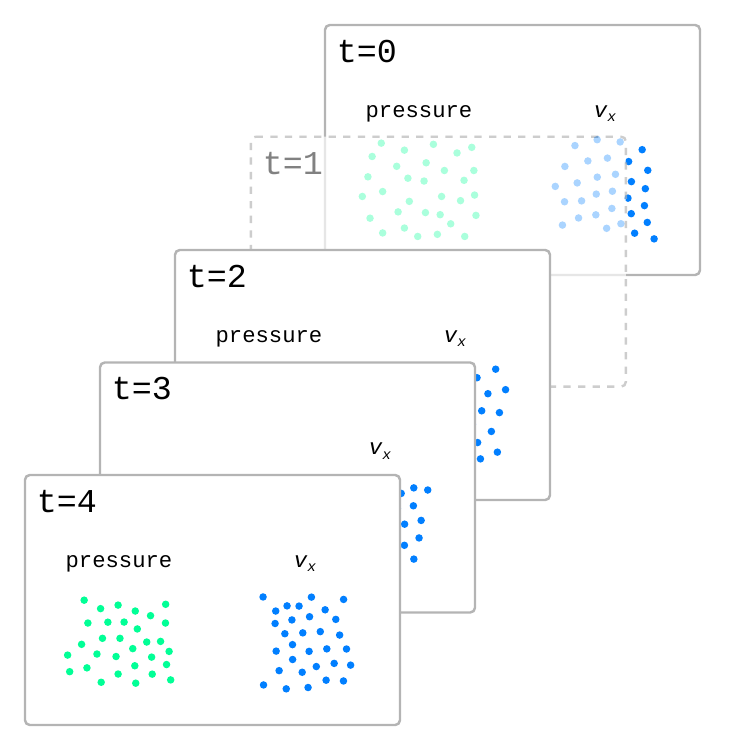}
\caption{Data points committed from different time steps are organized in time frames. Time frames can be non-uniformly distributed. Individual quantities can appear in a select subset, instead of all, of the time frames.}
\label{fig:tframe}
\end{figure}

To address this problem, MUI stores the collection of numerical results generated during each time step as a \textbf{frame}. Frames are indexed by their timestamps so different frames do not override each other. Technically, all data points being pushed in for a single physical quantity within a single time step are collectively stored in an instance of \replace{\code{mui::any\_storage},}{} MUI's dynamically typed data container. \replace{A frame is essentially a \code{std::map} between the quantity names and the actual \code{mui::any\_storage} instances}{A frame is a collection of mappings from quantity names to actual MUI data container objects}, while the frames themselves are again organized in a \replace{\code{std::map} using the time stamp as the key}{mapping where time stamps are used as indexing keys}. This sparse storage structure, as illustrated in Figure~\ref{fig:tframe}, allows efficient allocation of memory regardless of whether the time frames are equally distributed or not. It also allows each physical quantity to be selectively committed in a subset of all time frames.

\begin{figure}
\centering
\includegraphics[width=3.5in]{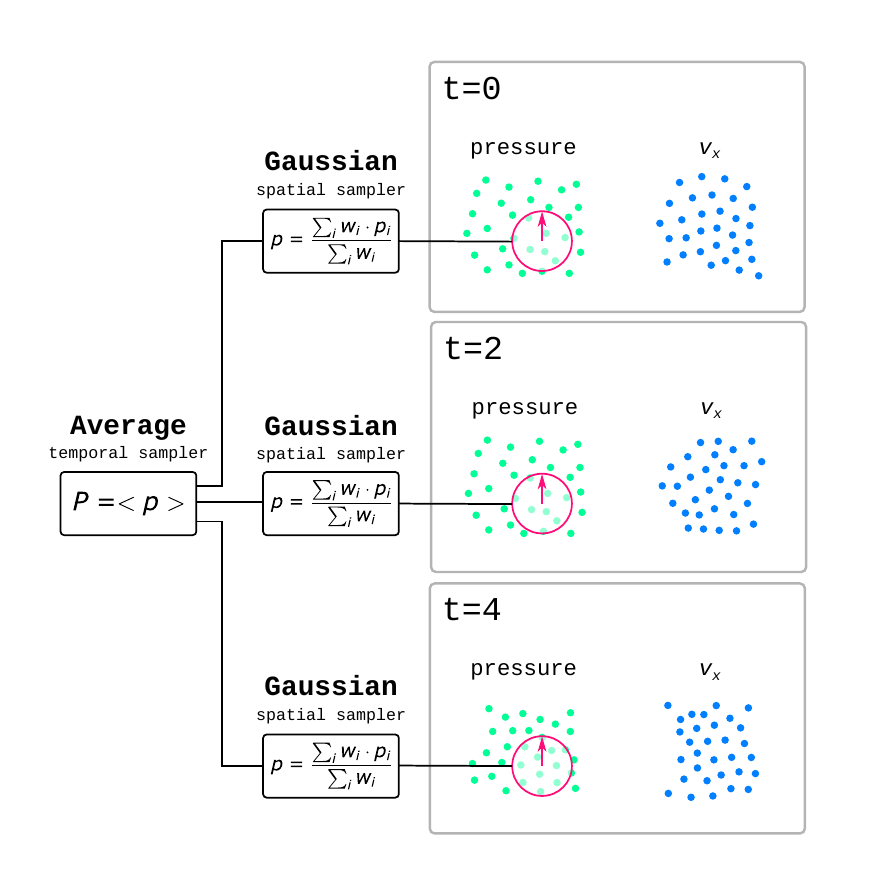}
\caption{Time samplers can interpolate or extrapolate the output of a spatial sampler over a range of time frames.}
\label{fig:tsampler}
\end{figure}

A set of \textit{time samplers} are also predefined in MUI. Time samplers work in essentially the same way as the data samplers, except for that they are one-dimensional along the time axis and use the output from a spatial sampler as the input. In Figure~\ref{fig:tsampler} we demonstrate the concept of a simple averaging sampler. It is also straightforward to implement more sophisticated time samplers with features such as filtering or prediction.

The memory allocation for frames is managed transparently by a buffering scheme. The deallocation, however, must be set up by user because it is impossible to predict whether a frame will be reused in the future. Utility methods are provided for \replace{the user}{users} to either explicitly request the disposal of time frames or to let MUI automatically discard frames that are older than a certain age in the simulation units. The default memory length is infinity so no frames will be freed automatically. In situations where batches of data points have to be moved between components of MUI, we use the C++11 \replace{\code{std::move}}{move} semantic to avoid duplicate memory allocation or copying.

\section{Parallel Communication}

\subsection{MPI Multiple-Program-Multiple-Data Setup}
\label{sec:MPIMPMD}

MUI uses MPI as the primary communication mechanism due to its portability, ubiquity, efficiency and compatibility with existing codes. The multiple-program-multiple-data (MPMD) mode of MPI is a natural fit for the purpose of concurrent coupling. \revised{In this mode, each solver is compiled and linked separately as independent executables but invoked simultaneously as a single MPI job using the MPMD syntax. An example of the MPMD launch syntax is given in Listing~\ref{list:mpmd}. Theoretically, users can launch an arbitrary amount of ranks for each of the solvers irrespective of the number of ranks spawned for its peer solvers.}

\begin{lstlisting}[caption=MPI MPMD launch syntax, label=list:mpmd]
mpirun -np n1 solver1 solver1_arguments : -np n2 solver2 solver2_arguments
\end{lstlisting}

\revised{The MUI inter-solver communication topology is established dynamically from the MPI runtime job configuration.} To ensure that this MPMD topology is only visible to MUI itself and hidden from the solver code, it is mandated by MUI that solvers should make no direct reference to the MPI predefined world communicator \mpicommworld for any of its own communications. Instead, a globally accessible \replace{variable of type \code{MPI\_Comm}}{MPI communicator, \textit{i.e.} a variable of type \code{MPI\_Comm},} should be defined to hold \replace{the \textit{global}}{a solver-specific global} communicator, which can be obtained from a MUI helper function call that effectively splits \revised{the MPI predefined world communicator} into subdomains using MPI application numbers. This is in fact one of the few modifications to the solvers that is ever dictated by MUI.

\begin{figure}
\centering
\includegraphics[width=3in]{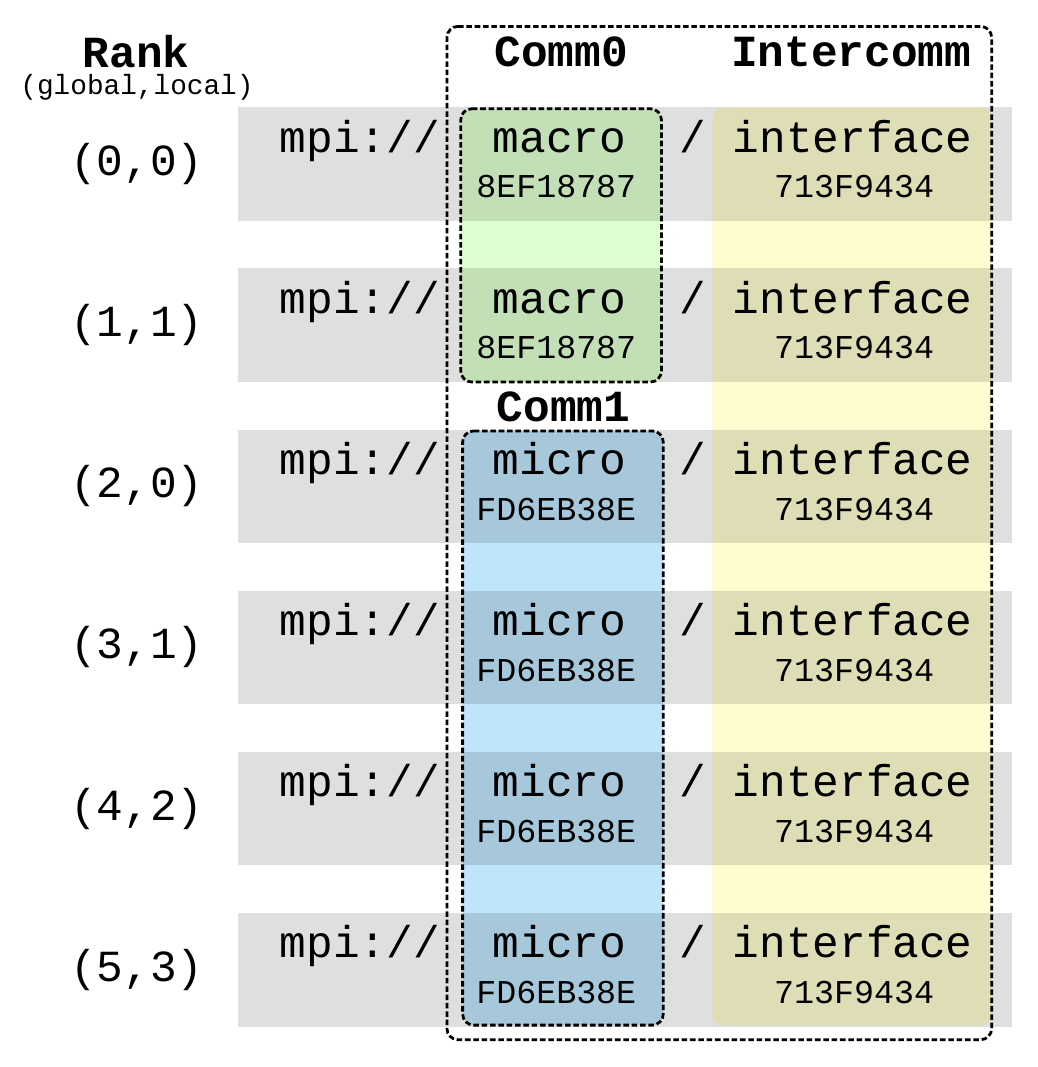}
\caption{\revised{In this example we demonstrate how an MPI job consisting of two macroscopic solver ranks and four microscopic solver ranks are partitioned according to the URI domain descriptor. The C++ \code{std::hash} functor can generate unique integer-valued hashes from the domain sub-string. MUI uses the hashes as the colors for splitting the MPI communicator. The hash value of the interface sub-string is then used to establish the inter-communicators that encompass both intra-communicators.}}
\label{fig:mpmd}
\end{figure}

\revised{In order to identify and connect MUI instances belonging to different domains, each solver instance needs to initialize MUI using a uniform resource identifier (URI) domain descriptor with the format \code{protocol://domain/interface}. The \code{protocol} field must always be \code{mpi} in the current MUI implementation. It indicates that the inter-solver communication manager sends and receives messages through MPI. Internally, the  communication managers are C++ objects allocated through an object factory mechanism, which would allow straightforward addition of new communication managers using alternative protocols such as TCP/IP or UNIX pipe. In our current communication scheme, the hash value of the \code{domain} sub-string is used as the \textit{color} for splitting the MPI built-in world communicators into smaller ones each containing a single physical domain. The \code{interface} sub-string is also hashed to generate a unique integer value for identifying different interface regions in the case of simulating multi-interface systems. The actual algorithm, as given in Algorithm~\ref{alg:mpimpmd}, makes use of the MPI \code{MPI\_COMM\_SPLIT} and \code{MPI\_INTERCOMM\_CREATE} method. Figure~\ref{fig:mpmd} demonstrates the setup process in a system consisting of two subdomains and an interface between the subdomains.}

\begin{algorithm}
\caption{MPI MPMD setup.}
\scriptsize
\ttfamily
\label{alg:mpimpmd}
\begin{algorithmic}
\Function{InitMpiMpmd}{URI}
  \LComment{Duplicate world communicator to avoid interfere with solver}
  \State World      $\gets$ \Call{MPI\_COMM\_DUP}{MPI\_COMM\_WORLD}
  \State GlobalSize $\gets$ \Call{MPI\_COMM\_SIZE}{World}
  \\
  \LComment{Parse URI string}
  \State DomainString, InterfaceString $\gets$ \Call{Parse}{URI}
  \State DomHash $\gets$ \Call{std::Hash}{DomainString}
  \State IfsHash $\gets$ \Call{std::Hash}{InterfaceString}
  \\
  \LComment{Create local and inter-communicators for solver}
  \State LocalDomain                 $\gets$ \Call{MPI\_COMM\_SPLIT}{World,DomHash}
  \State AllDomHash[0..GlobalSize-1] $\gets$ \Call{MPI\_ALLGATHER}{DomHash}
  \State AllIfsHash[0..GlobalSize-1] $\gets$ \Call{MPI\_ALLGATHER}{IfsHash}
  \State root $\gets$ $\min_i \mid$ AllDomHash[i] $\neq$ DomHash $\&$ AllIfsHash[i] $\equiv$ IfsHash
  \State RemoteDomain $\gets$ \Call{MPI\_INTERCOMM\_CREATE}{LocalDomain,0,World,root,IfsHash}
  \\
  \State \Return{LocalDomain,RemoteDomain}
\EndFunction
\end{algorithmic}
\end{algorithm}

\subsection{Asynchronous I/O and smart sending}
\label{sec:smartsend}

MUI assumes an asynchronous communication model because it can be difficult to find synchronization points between multiple heterogeneous solvers. Specifically, MPI collective methods are not used. Instead, MUI makes use of point-to-point non-blocking send and blocking receive methods. The send buffers are stored in a queue alongside with their corresponding MPI requests, and are freed upon completion of the communication. Whenever MUI finds itself in need of data (e.g., due to a fetch request), it continuously accepts incoming MPI messages while also testing for the completion of pending sends until the arrival of the needed data. This asynchronicity is encapsulated within MUI and is completely transparent to the solver. A non-blocking test method is also provided for advanced users to query the availability of data.

\begin{figure}
\centering
\includegraphics[width=3.5in]{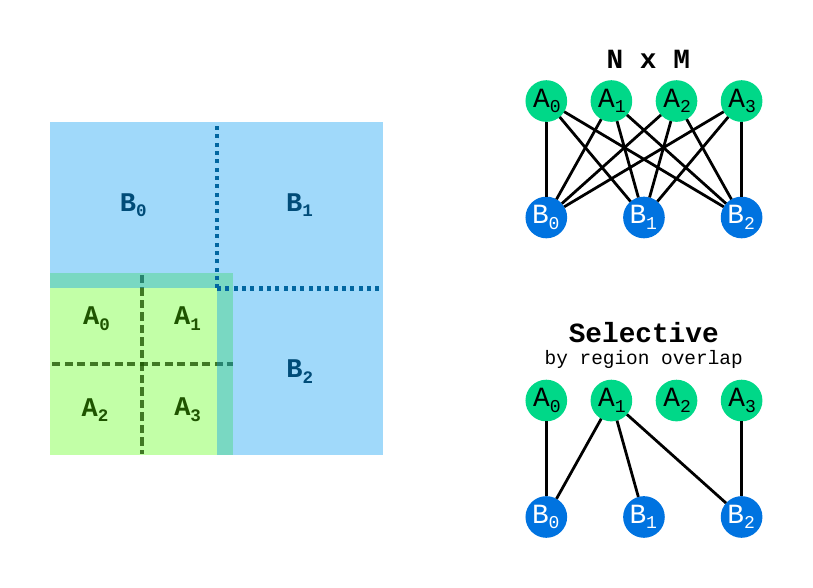}
\caption{An example illustrating the effect of smart sending. The green domain is handled by solver A and spatially decomposed between 4 MPI ranks, while the blue domain is handled by solver B using 3 MPI ranks. By checking for the overlap between the interface regions owned by the ranks, MUI can eliminate unnecessary data transfer between ranks such as $A_0-B_1$, $A_2-B_2$ and $A_3-B_0$ etc. }
\label{fig:smartsend}
\end{figure}

An optional \textit{smart sending} feature is also introduced to optimize the amount of MPI messages. It is a selective communication mechanism based on spatial overlap detection. Each solver instance can define two \textit{regions of interest}, \textit{i.e.} a fetch region and a push region, through a Boolean combination of geometric primitives such as spheres, cuboids and points. As illustrated in Figure~\ref{fig:smartsend}, the regions are broadcasted among all the processes so that the communication between a sender and a receiver whose regions of push/fetch have no overlap can be safely eliminated. In this way, the communication made by each MUI instance can be localized to a few peers who are truly in need of the data. To accommodate the case of moving boundaries, each region of interest is associated with a validity period. The smart sending feature can be safely ignored for convenience because both regions would default to the entire $\mathbb{R}^d$ with a validity period of infinity as a safety fall-back.

\section{Customizability}
\label{sec:custom}

MUI provides a vast customization space regarding communication content, spatial and temporal interpolation algorithm and communication pattern. In addition, MUI allows a number of its low-level traits to be parameterized at compile-time using a configuration class. A default configuration class \code{crunch} is shown in Listing~\ref{list:config}. The class serves as the last template argument of all MUI component classes and samplers, and is passed along during inheritance and member definition so users only need to specify it once when instantiating the MUI top-level object. It allows the tweaking of:
\begin{itemize}
  \item dimensionality of the physical space;
  \item precision of floating point numbers;
  \item integer width;
  \item time stamp type;
  \item type list (as mentioned in Section~\ref{sec:typing});
  \item debugging switch;
  \item exception handling \revised{behavior}.
\end{itemize}
Such static configuration mechanism can eliminate unnecessary runtime polymorphic overhead and also allows MUI to receive better performance optimization during the compilation phase.

\begin{lstlisting}[caption=The default configuration class for MUI, label=list:config]
struct crunch {
	static const int D = 3;

	using REAL       = double;
	using INT        = int64_t;
	using point_type = point<REAL,D>;
	using time_type  = REAL;

	static const bool DEBUG = false;

  template<typename... Args> struct type_list_t {};
	using type_list = type_list_t<int,double,float>;

	using EXCEPTION = exception_segv;
};
\end{lstlisting}


\section{Demonstration Examples}
\label{sec:benchmark}

\subsection{Couette flow: SPH-SPH coupling}
\label{sec:couette}

\replace{To demonstrate the real-world}{To give a concrete example of the} usage of MUI, we present a minimal-working-example (MWE) benchmark of concurrently coupled Smoothed Particle Hydrodynamics (SPH) simulation.

\begin{figure}
\centering
\includegraphics[width=3in]{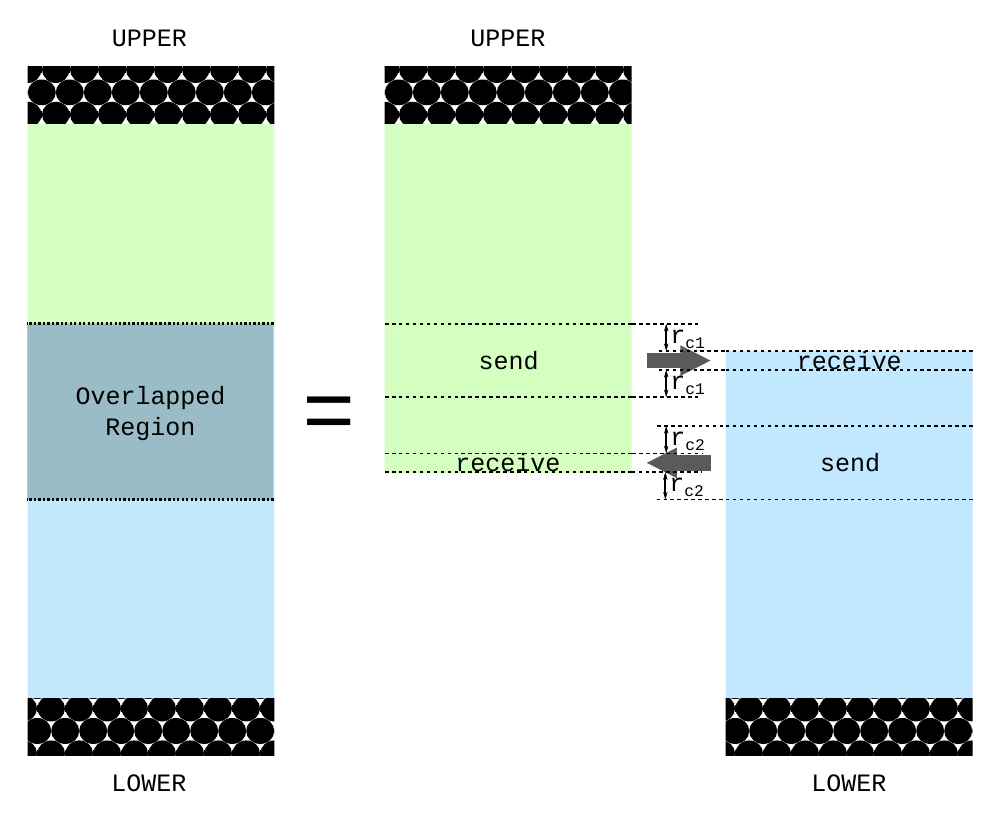}
\caption{\revised{The flow was simulated by two overlapping SPH domains. Each domain contains a send region and a receive region that are not overlapping with each other.}}
\label{fig:velocityscheme}
\end{figure}

The algorithm is based on a velocity coupling scheme described in Algorithm~\ref{alg:sphsphcoupling}. As illustrated in Figure~\ref{fig:velocityscheme}, the system was simulated using two overlapping SPH domains, \textit{i.e.} a lower one and an upper one, using either same or different resolutions. During each time step, the velocity of the SPH particles lying within the receiving part of the overlapped region is set as the average velocity of nearby particles from the other domain as interpolated using a SPH quintic interpolation sampler. We used LAMMPS \cite{plimpton2007lammps} as the baseline solver, and inserted only about 70 lines of code to implement the algorithm using MUI as given in Listing~\ref{SI-list:fixmui} in Support Information (SI).

\begin{algorithm}
\caption{SPH-SPH coupling scheme. The C++ code for the \code{Quintic} and \code{ExactTime} samplers are given in Listing~\ref{SI-list:chronoexact} and Listing~\ref{SI-list:quintic} in SI, respectively.}
\scriptsize
\ttfamily
\label{alg:sphsphcoupling}
\begin{algorithmic}
\For{$t$ = $0$:$\delta t$:$T_{total}$}
  \LComment{Push}
  \For{each particle $i$}
    \If{\Call{WithinSendRegion}{$i$}}
      \State \Call{MUI::Push}{"$v_x$",coord[$i$],vel$_x$[$i$]}
    \EndIf
  \EndFor
  \State \Call{MUI::Commit}{$t$}
  \LComment{Fetch}
  \For{each particle $i$}
    \If{\Call{WithinReceiveRegion}{$i$}}
      \State $S_{spatial}$ $\gets$ \Call{Quintic}{$r$,$h$}
      \State $S_{temporal}$ $\gets$ \Call{ExactTime}{$\varepsilon$}
      \State vel$_x$[$i$] $\gets$ \Call{MUI::Fetch}{"$v_x$",coord[$i$],$t$,$S_{spatial}$,$S_{temporal}$}
    \EndIf
  \EndFor
  \State \Call{MUI::Forget}{$t$}
\EndFor
\end{algorithmic}
\end{algorithm}

\begin{figure}
\centering
\includegraphics[width=3.5in]{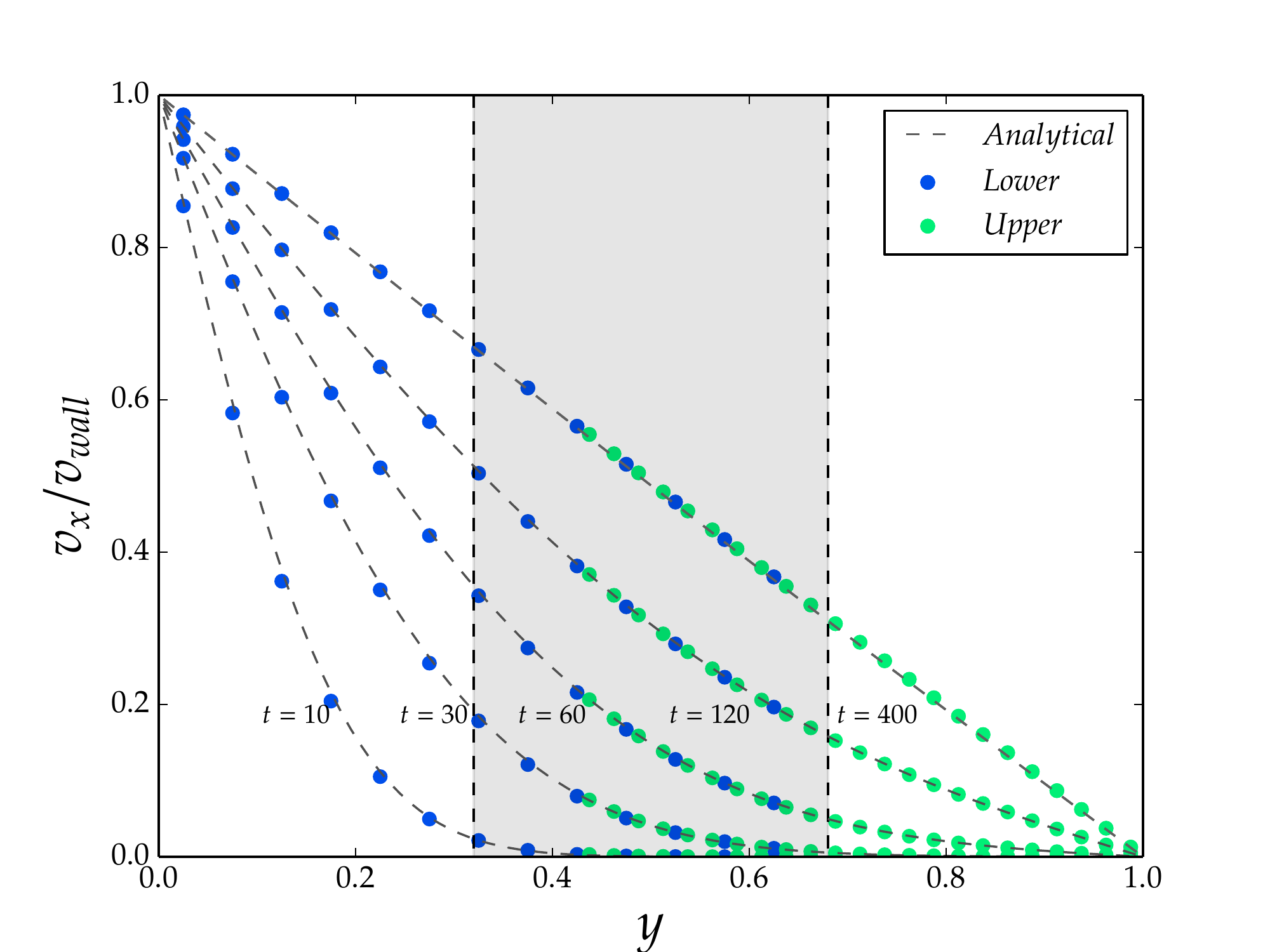}
\caption{Transient velocity profile of a Couette flow. The system was modeled by two SPH domains of different resolutions. Numerical solutions obtained from the lower and upper domains are plotted in blue and green, respectively. The analytic solutions are shown by dashed lines, while the interface region is indicated by the shaded box.}
\label{fig:poc}
\end{figure}

We then used the MUI-equipped LAMMPS to model a Couette flow by solving the Navier-Stokes equation. A system of a unit cube was simulated. The volumetric number densities of SPH particles were $20^3$ and $40^3$ for the lower and upper domains, respectively. As shown in Figure~\ref{fig:poc} the velocity profile obtained from the coupled simulation is consistent with the analytic solution.

\begin{figure}
  \centering
  \begin{subfigure}[t]{0.48\columnwidth}
    \centering
    \includegraphics[width=3in]{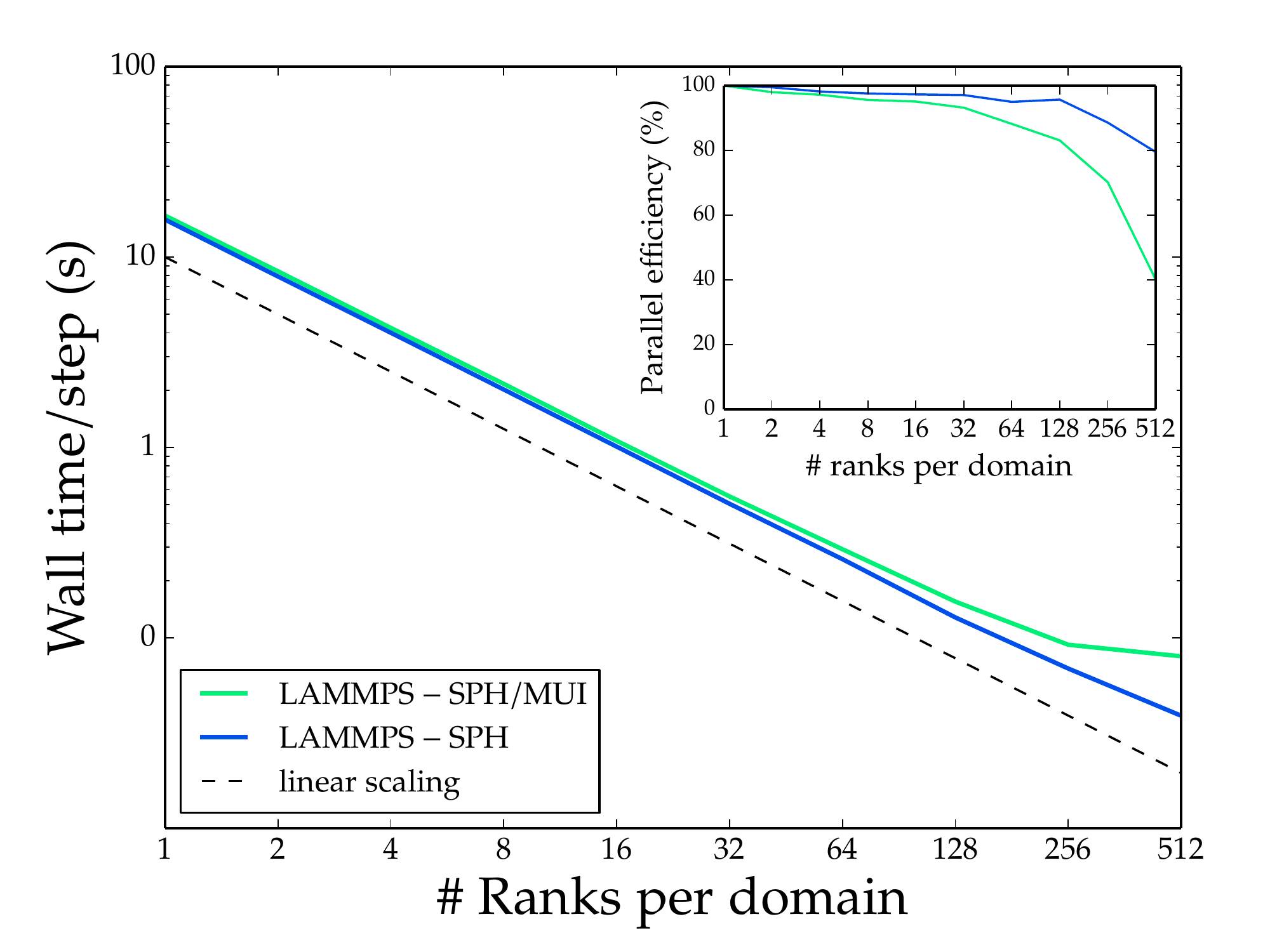}
    \caption{Breadth-first}
    \label{fig:scaling-breadth}
  \end{subfigure}
  \begin{subfigure}[t]{0.48\columnwidth}
    \centering
    \includegraphics[width=3in]{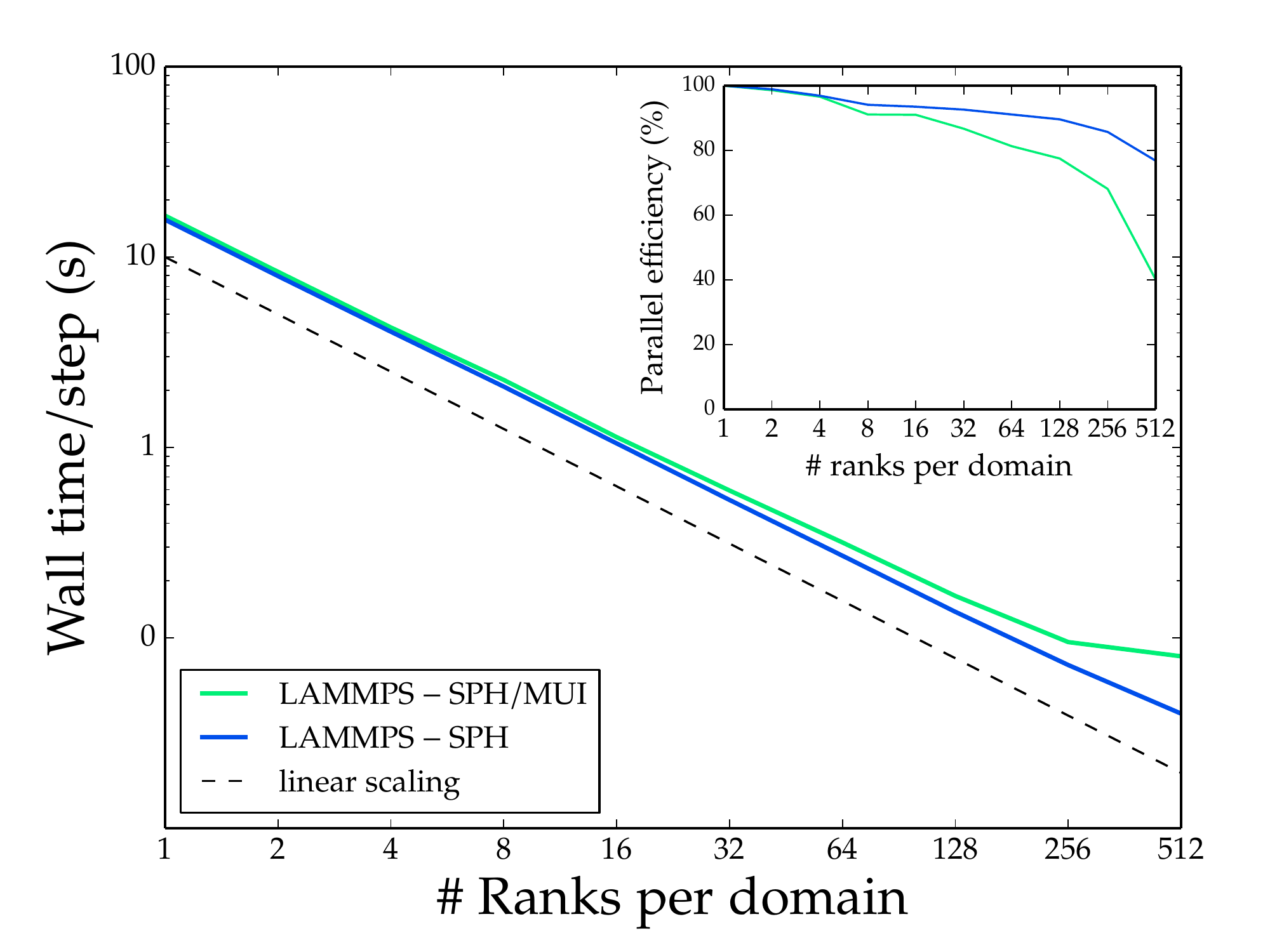}
    \caption{Depth-first}
    \label{fig:scaling-depth}
  \end{subfigure}
   \caption{\revised{Strong scaling performance comparison between the MUI-equipped and the original LAMMPS code.}}
  \label{fig:scaling}
\end{figure}

\begin{figure}
\centering
\includegraphics[width=3.5in]{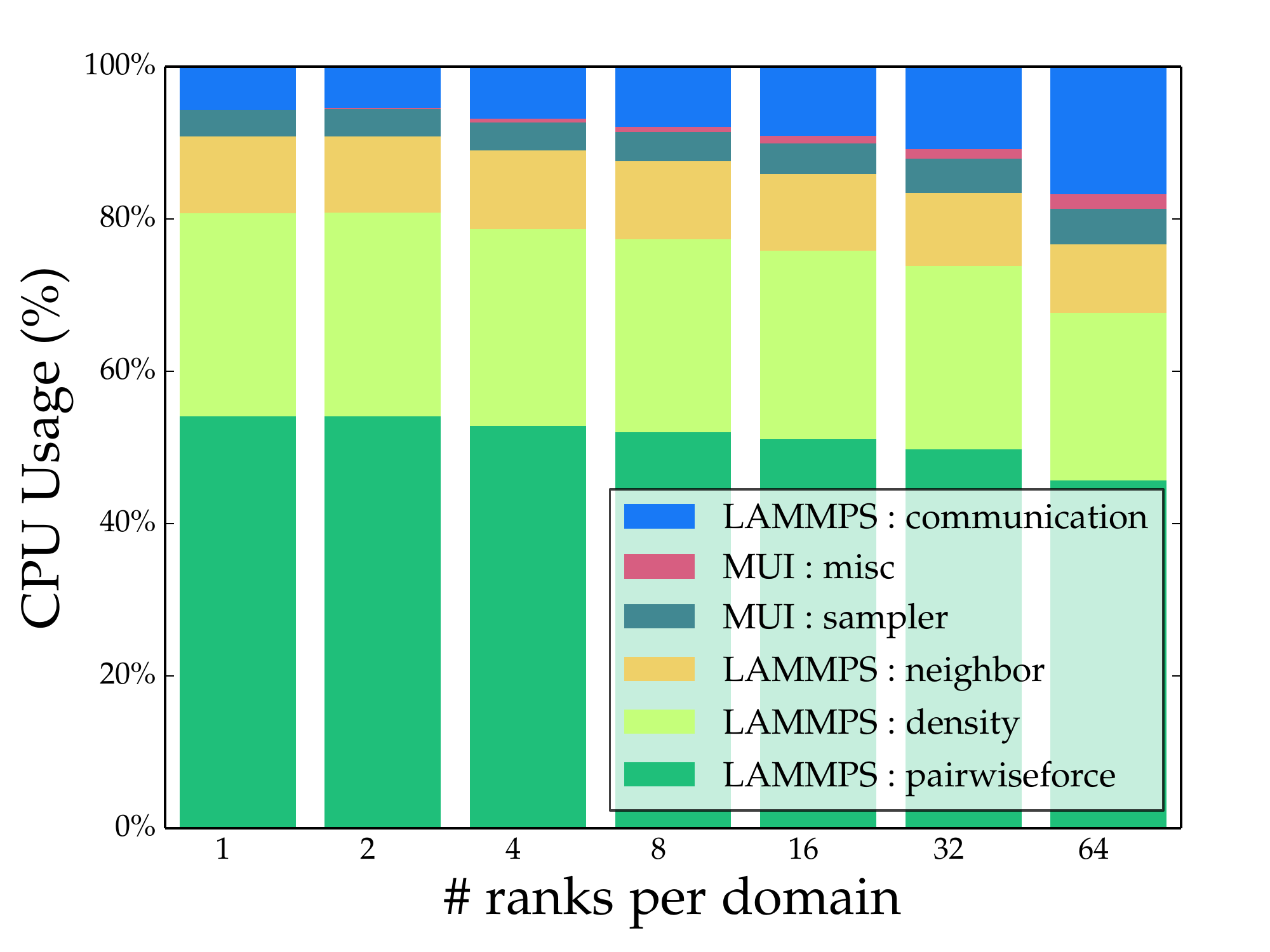}
\caption{\revised{Couette flow: a breakdown of the CPU time usage.}}
\label{fig:perf}
\end{figure}

\revised{
The performance and strong scalability of the MUI-equipped SPH solver were further characterized using a similar simulation setup. A system of size $6 \times 1 \times 6$ was simulated as a lower domain spanning from $0$ to $0.6125$ and an upper domain spanning from $0.3875$ to $1$. A number density of $40^3$ was used for both the lower and upper subdomains for the ease of performance evaluation. Each domain thus contained 1,382,400 fluid particles and 172,800 wall particles. The computation was performed on the Eos supercomputer using up to 512 ranks for each subdomain on a total of 128 nodes each containing 2 Intel Xeon E5-2670 CPUs at 2.6 GHz. The asynchronous progress engine feature of Cray MPI was enabled to better accommodate the communication pattern of MUI. Two different rank placement strategies, \textit{i.e.} breadth-first and depth-first, were used when increasing the amount of MPI ranks. With the breadth-first strategy, one MPI rank was spawned on each node until a maximum of 128 nodes were used. After that, the number of ranks per node was increased to further fill up the nodes until a total 1,024 ranks were spawned. With the depth-first strategy we first increased the number of ranks per node up to 8 before adding in more nodes. Baseline performance metrics were obtained by simulating the lower and upper domains independently using the original LAMMPS solver.

Figure~\ref{fig:scaling} visualizes the measured scalability and parallel efficiency of the MUI-equipped LAMMPS versus the baseline solver as obtained from the benchmark. Figure~\ref{fig:perf} presents a percentage breakdown of the CPU time spent in various parts of the solver at different levels of parallelism. In all cases, MUI consumes no more than 7\% of the total CPU time. Note that this includes both the sampling time which is actually part of the useful work and hence should not be counted as overhead.
}

\subsection{Soft Matter: SPH-DPD coupling}
\label{sec:softmatter}

\begin{figure}[t!]
\centering
\includegraphics[width=3.5in]{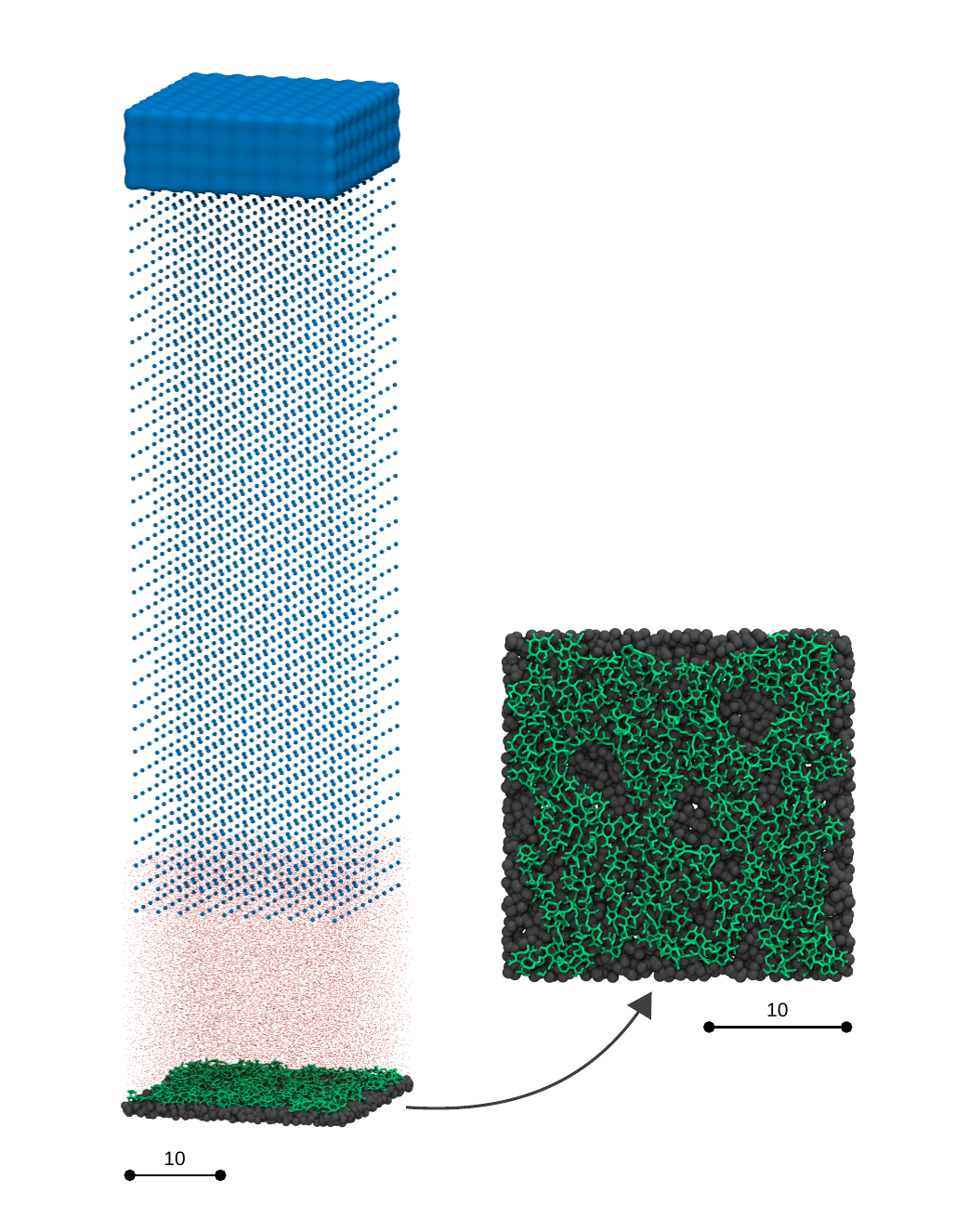}
\caption{A hybrid system consisting of an SPH upper domain and a DPD lower domain was modeled to study the hydrodynamical properties of dendrimer grafted surface. DPD wall particles, dendrimers, DPD solvent particles and SPH fluid particles are rendered in black, green, red and blue, respectively.}
\label{fig:grafted}
\end{figure}

Next we demonstrate a concurrently coupled deterministic/stochastic simulation using a similar coupling scheme. As illustrated in Figure~\ref{fig:grafted}, the flow between two parallel infinitely-large plates driven by a uniform body force was simulated. The upper plate corresponds to a simple no-slip boundary, while the lower plate is grafted by a hydrophobic fourth order binary dendrimers. We used a coupled simulation to investigate the effect of coating on the hydrodynamics of the system. \revised{All quantities/parameters mentioned in this simulation are in reduced DPD units.}

\revised{A system of size $40\times 110 \times 40$ was constructed using an upper domain and a lower domain. The parameters for setting up the simulation are listed in Table~\ref{table:sph-dpd-params}. A gravity of 0.001 in the x direction was imposed for both domains.}

\revised{The flow field in the upper domain with a simple no-slip boundary condition was simulated using the Smoothed Particle Hydrodynamics (SPH) method solving the Navier-Stokes equation. The SPH domain spanned from $y=20$ to $y=110$ and included a stationary upper wall lying between $y=100$ to $110$ which serves to enforce the no-slip boundary condition.}

\revised{The flow field in the lower domain was simulated using Dissipative Particle Dynamics (DPD) solving Newton's equation of motion in stochastic form. The DPD domain spanned from $y=-1$ to $y=28$, and included the solvent and a stationary wall lying between $y=-1$ to $y=0$ with its upper surface grafted by 160 fourth order binary dendrimers. The surface converage of the dendrimers was 70\%. The viscosity of the DPD solvent was measured as 3.72.}

\begin{figure}
\centering
\includegraphics[width=3.5in]{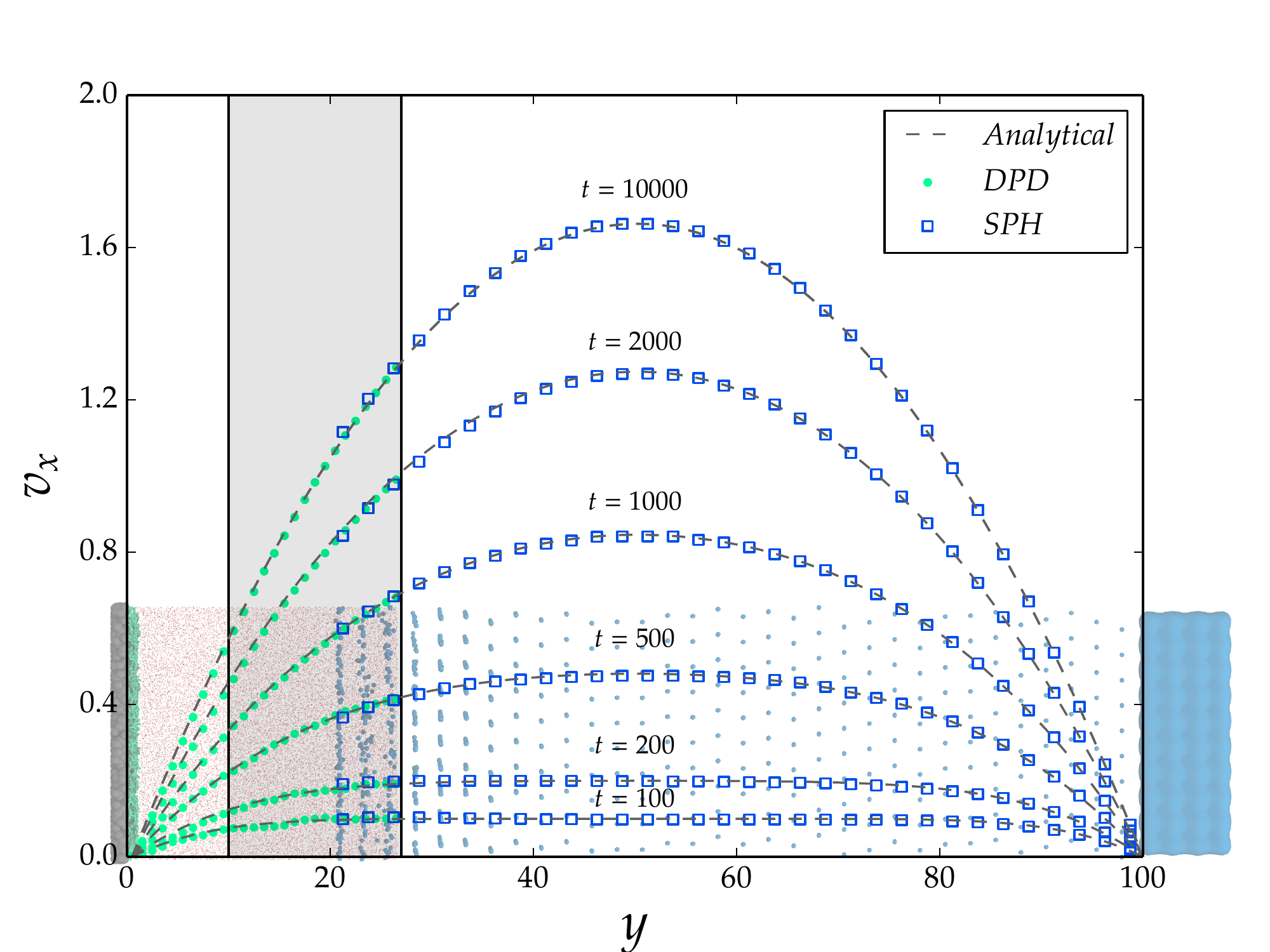}
\caption{The velocity profile derived from the coupled SPH/DPD simulation converges to that of a Poiseuille flow. SPH results, DPD results and the analytic solutions are plotted in blue dots, green dots and dashed line, respectively. The background picture indicates the system composition at the corresponding $y$ position.}
\label{fig:grafted_prof}
\end{figure}

The \revised{MUI-based} coupling scheme is similar to that use in the previous SPH-SPH simulation as described in Section~\ref{sec:couette}. However, the SPH solver assumes a time step size which is 50 times that of the DPD solver. Accordingly, as demonstrated in Algorithm~\ref{alg:sphdpdcoupling}, the SPH domain samples the average velocity of the DPD domain over the last 50 frames to smooth out the randomness, while the DPD domain always samples the latest time frame sent from the SPH domain. The velocity profile converged to that of a Poiseuille flow after 10000 DPD units. As shown in Figure~\ref{fig:grafted_prof} the effective channel width is reduced by the hydrophobic coating by about 0.5 DPD unit.

\revised{The simulation was done on a workstation with two quad-core Intel Xeon E5-2643 CPUs running at 3.30GHz as well as four nVidia GeForce GTX TITAN GPUs each with 2688 cores. The DPD simulation was done on the four GPUs using the \USERMESO package~\cite{tang2013accelerating}, while the SPH code ran on a single CPU core using the same LAMMPS SPH solver as mentioned in the previous example. This processor resource allocation was based on the obvervation that simulating the DPD domain is orders of magnitude more expensive than simulating the SPH domain.}

\begin{table}
\caption{Soft matter: parameters for the SPH-DPD simulation}
\label{table:sph-dpd-params}
\centering
\begin{tabular}{cclcccl}
\toprule
\multicolumn{3}{c}{DPD} && \multicolumn{3}{c}{SPH} \\
arg & val & \multicolumn{1}{c}{\revised{description}} && arg & val & \multicolumn{1}{c}{\revised{description}} \\
\cmidrule{1-3} \cmidrule{5-7}
$N_p$      & $84,375$ & number of particles     &&  $N_p$      & $4,000$ & number of particles \\
$\rho$     & $5$      & particle number density &&  $\rho_N$   & $0.064$ & particle number density \\
$\gamma$   & $4.5$    & noise level             &&  $\rho_m$   & $5$     & mass density \\
$\sigma$   & $3.0$    & dissipation             &&  $\eta$     & $3.72$  & viscosity \\
$\delta t$ & $0.01$   & time step               &&  $\delta t$ & $0.5$   & time step \\
$r_c$      & $1.0$    & cutoff distance         &&  $r_c$      & $10$    & cutoff distance \\
$k_B T$    & $1.0$    & temperature level       &&  $c_s$      & $1.5$   & speed of sound\\
$g$    & $0.001$      & body force              &&  $g$        & $0.001$ & body force \\
\multicolumn{2}{c}{$w_D = w_C^{0.5}$} & weight function && & \\
\bottomrule
\end{tabular}
\end{table}

\begin{table}
\centering
\caption{Repulsive force constants $a_{ij}$ for DPD}
\label{table:aij}
\begin{tabular}{cccc}
\toprule
           &   wall & dendrimer & solvent \\
wall       &     15 &        15 &    15   \\
dendrimer  &     15 &        15 &    75   \\
solvent    &     15 &        75 &    15   \\
\bottomrule
\end{tabular}
\end{table}

\begin{algorithm}
\caption{SPH-DPD coupling scheme. The C++ code for the \code{SumOver} sampler is given in Listing~\ref{SI-list:chronosum} in SI.}
\scriptsize
\ttfamily
\label{alg:sphdpdcoupling}
\begin{algorithmic}
\LComment{SPH domain}\\
\For{\hlcode{$t$ = $0$:$50\delta t$:$T_{total}$}}
  \LComment{Push}
  \For{each particle $i$}
    \If{\Call{WithinSendRegion}{$i$}}
      \State \Call{MUI::Push}{"$v_x$",coord[$i$],vel$_x$[$i$]}
    \EndIf
  \EndFor
  \State \Call{MUI::Commit}{$t$}
  \LComment{Fetch}
  \For{each particle $i$}
    \If{\Call{WithinReceiveRegion}{$i$}}
      \State $S_{spatial}$ $\gets$ \Call{Quintic}{$r_{DPD}$,$h_{DPD}$}
      \State $S_{temporal}$ $\gets$ \hlcode{\Call{SumOver}{$50\delta t$}}
      \State vel$_x$[$i$] $\gets$ \Call{MUI::Fetch}{"$v_x$",coord[$i$],$t$,$S_{spatial}$,$S_{temporal}$}
    \EndIf
  \EndFor
  \State \Call{MUI::Forget}{$t$}
\EndFor
\\
\LComment{DPD domain}
\For{\hlcode{$t$ = $0$:$\delta t$:$T_{total}$}}
  \LComment{Push}
  \For{each particle $i$}
    \If{\Call{WithinSendRegion}{$i$}}
      \State \Call{MUI::Push}{"$v_x$",coord[$i$],vel$_x$[$i$]}
    \EndIf
  \EndFor
  \State \Call{MUI::Commit}{$t$}
  \LComment{Fetch}
  \State \hlcode{$t_{\mathrm{SPH}}$ $\gets$ \Call{Floor}{$t$,$50\delta t$}}
  \For{each particle $i$}
    \If{\Call{WithinReceiveRegion}{$i$}}
      \State $S_{spatial}$ $\gets$ \Call{Quintic}{$r_{SPH}$,$h_{SPH}$}
      \State $S_{temporal}$ $\gets$ \hlcode{\Call{ExactTime}{$\varepsilon$}}
      \State vel$_x$[$i$] $\gets$ \Call{MUI::Fetch}{"$v_x$",coord[$i$],\hlcode{$t_{\mathrm{SPH}}$},$S_{spatial}$,$S_{temporal}$}
    \EndIf
  \EndFor
  \If {\Call{Mod}{$t$,$50\delta t$} $=$ $0$}
    \State \Call{MUI::Forget}{$t-50\delta t$}
  \EndIf
\EndFor
\end{algorithmic}
\end{algorithm}

\subsection{Conjugate Heat Transfer}
\label{sec:heattransfer}

We further demonstrate a coupled Eulerian/Lagrangian simulation of the cooling process of a heating cylinder immersed in a channel flow using the energy-conserving Dissipative Particle Dynamics (eDPD) method \cite{li2014edpd} and the finite element method (FEM). The eDPD model is an extension to the classical DPD model \cite{Hoogerbrugge1992DPD,espanol1995dpd} with explicit temperature and heat transferring terms. The FEM solver can solve the time-dependent heat equation
\begin{align*}
\frac{\partial T}{\partial t} = -\alpha \Delta T + f
\end{align*}
where $\alpha$ is the thermal diffusivity and $f = f_0 + f_{\mathrm{interface}}$ the heat source.

\begin{figure}
\centering
\includegraphics[width=3.0in]{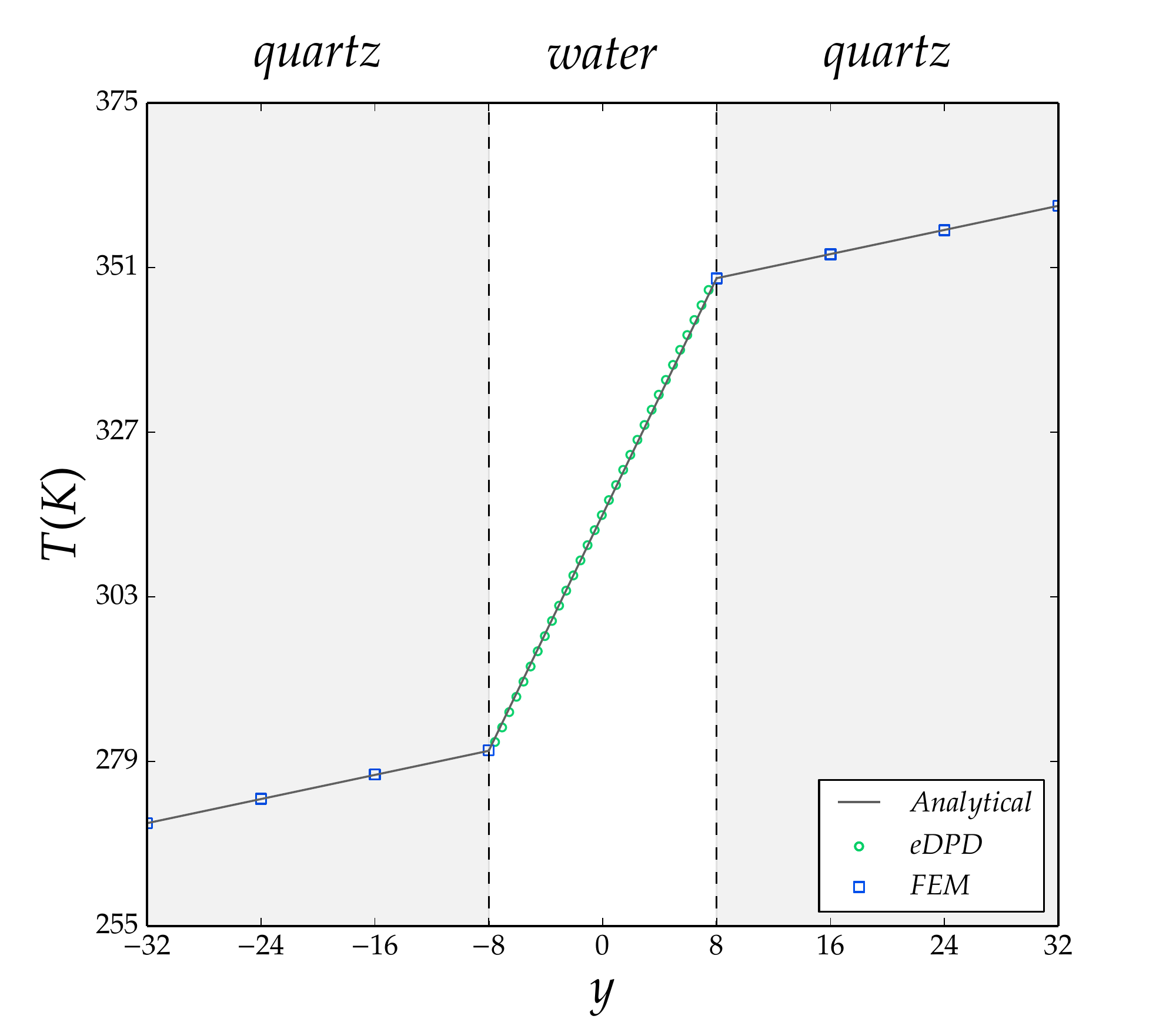}
\caption{Heat conduction: temperature profile obtain for a quartz-water-quartz tri-layer system with the FEM solver handling the solid domain and the eDPD solver handling the fluid domain.}
\label{fig:femvalidate}
\end{figure}

The coupling scheme is shown in Algorithm~\ref{alg:edpdfemcoupling}. The FEM domain pushes the temperature of boundary vertices, with which the eDPD solver calculates the heat flux generated by each particles surrounding the cylinder. The eDPD solver then pushes the flux data back into the FEM solver. The FEM solver averages the fluxes computed by eDPD and assigns the result to boundary vertices based on a Voronoi diagram of the vertices. The heat flux value is used as the Neumann boundary condition required in solving the time-dependent Poisson equation. The accuracy of the scheme was validated by solving for the temperature profile in a quartz-water-quartz system whose left and right boundaries were fixed at $270K$ and $360K$ as shown in Figure~\ref{fig:femvalidate}. The thermal diffusivities of water and quartz were assumed to be constant at $0.143\times 10^6 m^2/s$ \cite{blumm2007characterization} and $1.4\times 10^6 m^2/s$ \cite{gibert2009quartz} over the temperate range, respectively, while the interfacial thermal diffusivity was chosen as the arithmetic mean between the two values.

\begin{algorithm}
\caption{eDPD-FEM coupling scheme. The C++ code for the \code{VoronoiMean} and \code{Linear} samplers are given in Listing~\ref{SI-list:samplervoronoi} and Listing~\ref{SI-list:samplerlinear} in SI, respectively.}
\scriptsize
\ttfamily
\label{alg:edpdfemcoupling}
\begin{algorithmic}
\LComment{eDPD domain}
\For{$t$ = $0$:$\delta t$:$T_{total}$}
  \LComment{Push}
  \State $t_{\mathrm{FEM}}$ $\gets$ \Call{Floor}{$t$,$10\delta t$}
  \For{each particle $i$}
    \If{\Call{WithinCutoffOfCylinder}{$i$}}
      \State $S_{spatial}$ $\gets$ \hlcode{\Call{Linear}{$h_{max}$}}
      \State $S_{temporal}$ $\gets$ \Call{ExactTime}{$\varepsilon$}
      \State $T_{wall}$ $\gets$ \Call{MUI::Fetch}{"$T$",coord[$i$],$t_{\mathrm{FEM}}$,$S_{spatial}$,$S_{temporal}$}
      \State $q$ $\gets$ \Call{PerParticleHeatFlux}{T[$i$],$T_{wall}$}
      \State \Call{MUI::Push}{"$q$",coord[$i$],$-q / C_v$}
    \EndIf
  \EndFor
  \State \Call{MUI::Commit}{$t$}
  \If {\Call{Mod}{$t$,$10\delta t$} $=$ $0$}
    \State \Call{MUI::Forget}{$t-10\delta t$}
  \EndIf
\EndFor
\\
\LComment{FEM domain}\\
\For{$t$ = $0$:$10\delta t$:$T_{total}$}
  \LComment{Push}
  \For{each boundary vertex $i$}
    \State \Call{MUI::Push}{"$T$",coord[$i$],T[$i$]}
  \EndFor
  \State \Call{MUI::Commit}{$t$}
  \LComment{Fetch}
  \For{each boundary vertex $i$}
    \State $S_{spatial}$ $\gets$ \hlcode{\Call{VoronoiMean}{Vertices}}
    \State $S_{temporal}$ $\gets$ \Call{MeanOver}{$10\delta t$}
    \State f$_{interface}$[$i$] $\gets$ \Call{MUI::Fetch}{"$q$",coord[$i$],$t$,$S_{spatial}$,$S_{temporal}$}
  \EndFor
  \State \Call{MUI::Forget}{$t$}
  \State \Call{SolveForNextStep}{}
\EndFor
\end{algorithmic}
\end{algorithm}

\begin{figure*}
\centering
\includegraphics[width=7in]{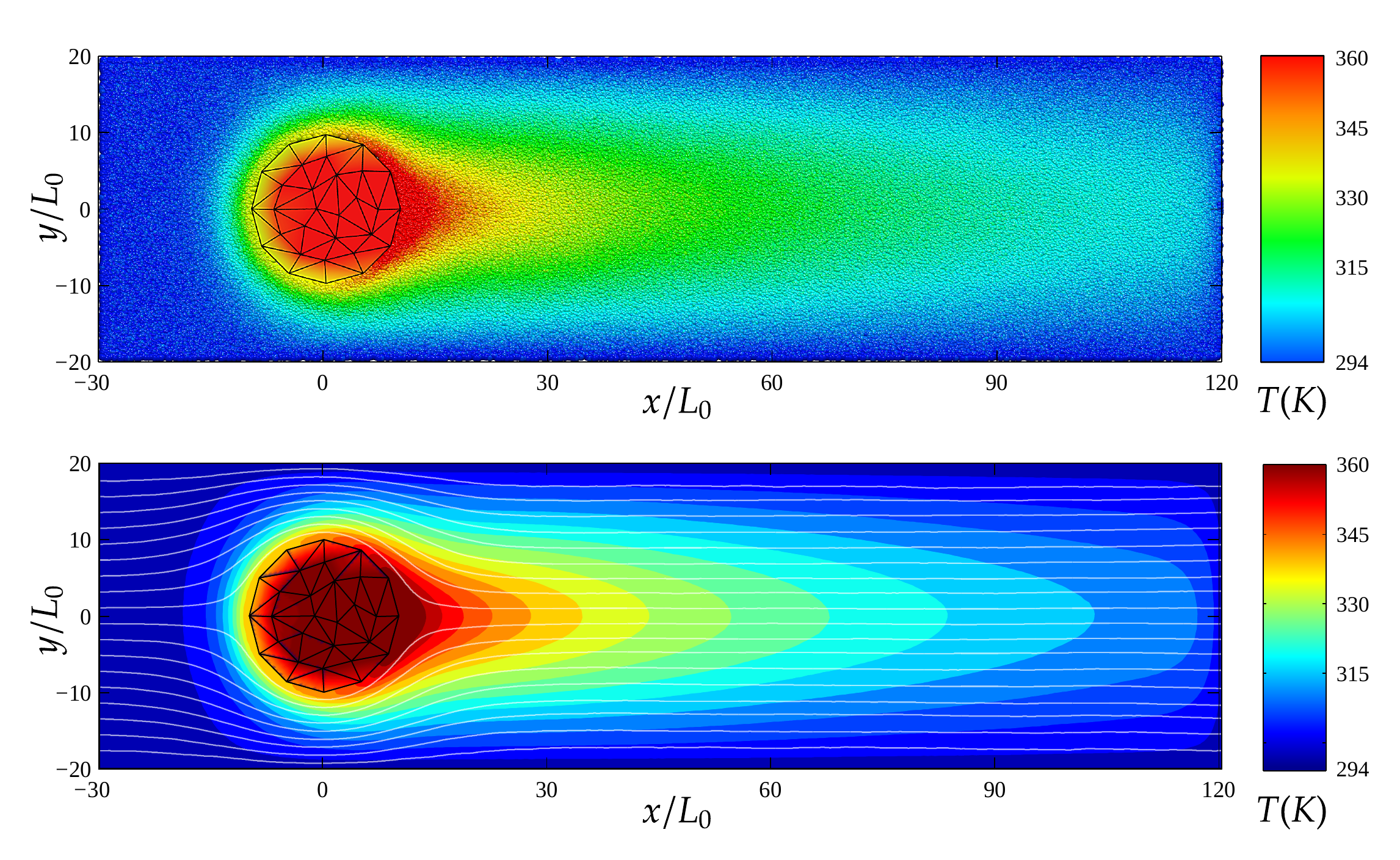}
\caption{Conjugate heat transfer: a snapshot of the hybrid particle/mesh structure of the domain at steady state is shown in the upper plot; streamlines and the temperature field in the system at steady state are visualized in the lower plot by white lines and color-mapped contours, respectively.}
\label{fig:cooling}
\end{figure*}

A system composed of a 3D fluid domain filled with water, a 2D solid domain and a fluid-solid interface was then simulated using the validated scheme. The entire domain is periodic in x and z direction, but is bounded by a pair of no-slip infinite walls of constant temperature in the y direction. The fluid domain and the walls are simulated using eDPD, while the solid domain is solved using FEM. Parameters and scaling factors used to set up the eDPD and FEM calculations are given in Table~\ref{table:edpd-fem-params}. The simulation result is shown in Figure~\ref{fig:cooling}. The Reynolds number is defined by $Re = (v_{max} D)/\nu = 1.97$ where $v_{max}=0.65 L_0 / \tau$ is the maximum inlet velocity, $\nu=6.62$ the kinematic viscosity and $D=20.0 L_0$ the diameter of the cylinder. 

We obtained a smooth temperature transition in the hybrid domain using MUI and the aforementioned coupling scheme. The example demonstrates the capability of MUI in coupling two different fields, \textit{e.g.} a flow field and a thermal field. The method can facilitate the study of inhomogeneous coolants, \textit{e.g.} colloidal suspensions, thanks to the flexibility brought about by the particle method.

\revised{The simulation was performed on a workstation with two hexa-core Intel Xeon E5-2630L CPUs running at 2.0GHz. The eDPD simulation occupied 11 CPU cores using our customized LAMMPS code. The FEM solver ran on a single CPU core. This computational resource configuration was based the relative computational cost of the two codes.}

\begin{table}
\caption{Conjugate heat transfer: parameters for the eDPD-FEM simulation of immersed heating cylinder are taken from Ref~\cite{li2014edpd}.}
\label{table:edpd-fem-params}
\centering
\begin{tabular}{lcclc}
\toprule
\multicolumn{2}{l}{length scale}      && \multicolumn{2}{l}{$L_0 = 11 nm$} \\
\multicolumn{2}{l}{time scale}        && \multicolumn{2}{l}{$\tau = 0.935 ns$} \\
\multicolumn{2}{l}{temperature scale} && \multicolumn{2}{l}{$T_0 = 300 K$} \\
\multicolumn{2}{l}{mass scale}        && \multicolumn{2}{l}{$m_0 = 3.32\times 10^{-22} kg$} \\
\midrule
\multicolumn{2}{c}{eDPD} && \multicolumn{2}{c}{FEM} \\
arg & val && arg & val \\
\cmidrule{1-2} \cmidrule{4-5}
$\alpha$   & $1.43\times 10^{-7} m^2/s$  &&  $\alpha$ & $1.43\times 10^{-7} m^2/s$ \\
$\delta t$ & $0.0125 \tau$  &&  $\delta t$ & $0.125 \tau$ \\
$\nu$      & $8.57\times 10^{-7} m^2\cdot s^{-1}$         &&  $f_0$      & $0.004 T_0/\tau$\\
& && space & $P_1$\\
\bottomrule
\end{tabular}
\end{table}

\section{Related work}
\label{sec:related}

Many software tools and frameworks have been proposed for carrying out concurrently coupled multiscale simulation. However, MUI differentiates itself from the existing ones by its ease of use, universaility, code reusability and out-of-the-box parallel communication capability.

The majority of existing frameworks focus on the coupling between solvers that deal with PDEs using grid/mesh-based methods. The Core Component Architecture (CCA) specification~\cite{Allan2002CCA,Lefantzi2003CCA,mcinnes2006PDE} appears to be the most widely adopted standard for this types of coupling with conforming implementations such as Uintah~\cite{uintah2000,Parker2006Uintah,Berzins2010uintah}, CCAT~\cite{bramley2000CCAT}, SCIRun2~\cite{zhang2004scirun2} and so on~\cite{gannon2004component,li2013CCAmegneto}. The CCA framework invokes individual solvers as components at runtime and let components register interface functions, called \textit{ports}, that are to be called by other components for information exchange. Different from MUI, the CCA specification does not specify the input and output arguments of the ports, and it is up to the solvers to determine the content of communication. The CCA core specification also does not define a mechanism for inter-solver communication outside of the shared memory space of a single process. \revised{Compared to MUI, the interfaces exposed by CCA-based frameworks typically require more engineering effort to cover lower level program activities such as inter-rank communication protocol handling, load balancing and memory management.} There exist also coupling frameworks that require more extensive modification of existing solvers. Solving algorithms are encapsulated as classes that expose a given set of interfaces in the Kratos~\cite{Dadvand2013Kratos} framework. The DDEMA~\cite{michopoulos2003DDEMA} framework focuses on PDE solvers using mesh-based methods and casts software packages into an \textit{actor} design pattern. \revised{MUI, on the contrary, does not require code refactoring and only requires the insertion of a few lines of pushing/fetching code.}

In addition, various specialized coupling frameworks have been proposed. \revised{For example, the PPM library~\cite{sbalzarini2006ppm} represents one of the major efforts in carrying out simulations with a unified particle and mesh representation.} \revised{The triple-decker algorithm can solve multiscale flow fields by coupling the Molecular Dynamics method, the Dissipative Particle Dynamics method and the incompressible Navier-Stokes equations~\cite{fedosov2009triple}. It is essentially a mathematical framework that formulates the way in which information should be exchanged at subdomain boundaries. In fact, this algorithm can be implemented with MUI in a straightforward manner.} The Macro-Micro-Coupling~\cite{neumann2013mamico,Neumann2014hybrid} framework can solve coupling problems between macroscopic models and microscopic models. The MCI~\cite{grinberg2011mci} framework is able to couple massively parallel spectral-element simulations and particle-based simulations in a highly efficient and scalable manner. The MUPHY~\cite{Bernaschi2009MUPHY} framework couples Lattice-Boltzmann method with molecular dynamics simulation. The MUSE~\cite{zwart2009MUSE,zwart2013multiphysics} framework is specialized in astrophysics simulation. \revised{In contrast, MUI is able to accomodate any method because it allows the encoding of solver-/method-specific information as data points of arbitrary custom value.} It can facilitate the construction of a plug-and-play pool of any combination of particle-based and continuum-based solvers for solving multiscale problems without code rewriting. To the best of our knowledge, There is currently no other project that can achieve such level of generality.


\section{Conclusion}

In this paper we presented the Multiscale Universal Interface library as a generalized approach of coupling heterogeneous solver codes to perform multi-physics and multiscale simulations. The library assumes a solver/scheme-agnostic approach in order to accommodate as many numerical methods and coupling schemes as possible, while still maintains a simple and straightforward programming interface. The \textit{data sampler} concept is the key enabling technique for this flexible framework of data interpretation. The library employs techniques such as dynamic typing, MPI MPMD execution, asynchronous I/O, generic programming and template metaprogramming to improve both performance and flexibility. Benchmarks demonstrate that the library delivers excellent parallel efficiency and can be adopted easily for coupling heterogeneous simulations. \revised{It is intended to be a development tool that can be used for fast implementation of abstract mathematical frameworks and coupling methodologies.}


\section*{Appendix}
\subsection*{Language compatibility}
\label{sec:lang}

\revised{The main body of MUI is written in C++11, the version of ISO standard C++ ratified in 2011. Due to the extensive usage of template programming techniques and C++11 newly-introduced features, the code has to be compiled with more recent versions of C++ compilers that have complete C++11 support. Table~\ref{table:compiler} summarizes our experimental result on the versions of common compilers which were able to build MUI-enabled solvers. The list is conservative in that older compilers may also work depending on the specific situation. To compile with C++11, a single compiler option, \textit{e.g.} \code{-std=c++11} for GCC, has to be added into the compiler command line. A wrapper for MUI is needed for projects developed in other languages. A sample C wrapper and a sample Fortran wrapper are included in MUI, while adaptations to other languages can be made relatively easily.}

\begin{table}
\caption{\revised{Compilers compatibility of MUI. For those not currently supporting MUI, full C++11 feature support was also announced and will be available within the near future.}}
\label{table:compiler}
\centering
\begin{tabular}{llll}
\toprule
 Compiler & Maintainer & Version & Flag \\
\midrule
GCC        & GNU        & 4.8.3     & \code{-std=c++11} \\
Clang      & Apple      & 3.5.0     & \code{-std=c++11} \\
Intel C++  & Intel      & 15.0      & \code{-std=c++11},~\code{/Qstd=c++11} \\
NVCC       & nVIDIA     & 7.0       & \code{-std=c++11} \\
Visual C++ & Microsoft  & -         & \\
XL C++     & IBM        & -         & \\
PGCC       & PGI        & -         & \\
\bottomrule
\end{tabular}
\end{table}

\section*{Acknowledgment}

This work was supported by the Department of Energy (DoE) Collaboratory on Mathematics for Mesoscopic Modeling of Materials (CM4) and the Air Force Office of Scientific Research (FA9550-12-1-0463). Simulations were carried out at the Oak Ridge Leadership Computing Facility through the Director Discretion project BIP102 and the Innovative and Novel Computational Impact on Theory and Experiment program BIP118. Particle simulation results were visualized with VMD \cite{HUMP96,STON1998}.

\section*{Reference}

\bibliography{coupling}
\bibliographystyle{cpc}

\end{document}